\documentclass[12pt]{report}
\usepackage[utf8]{inputenc}
\usepackage{charter}
\usepackage{fullpage}
\usepackage[hyphens]{url}
\usepackage{hyperref}
\hypersetup{hidelinks}
\usepackage{graphicx}
\usepackage{lipsum}
\usepackage[total={6.5in,9in},top=1in,headsep=0.1in,headheight=1in]{geometry}

\usepackage{longtable}

\usepackage{amssymb}
\usepackage{amsmath}
\usepackage{xcolor}
\usepackage{verbatim}
\usepackage{multirow}
\usepackage{tiffany}
\usepackage{enumitem}

\usepackage[backend=biber, sorting=none]{biblatex}
\usepackage{xspace}
\newcommand{\degree}{\ensuremath{^\circ}\xspace}

\newcommand{\htp}{\ensuremath{\mathrm{H}_2^+}\xspace}
\newcommand{\nuebar}{\ensuremath{\bar{\nu}_e}\xspace}
\newcommand{\nue}{\ensuremath{\nu_e}\xspace}

\newcommand{\numu}{\ensuremath{\nu_{\mu}}\xspace}

\newcommand{\nutau}{\ensuremath{\nu_{\tau}}\xspace}
\newcommand{\htct}{\ensuremath{\mathrm{H2C2}}\xspace}

\newcommand\snowmass{
\begin{center}
  \rule[-0.2in]{\hsize}{0.01in}\\
  \rule{\hsize}{0.01in}\\
  Submitted to the Proceedings of the US Community Study\\ 
  on the Future of Particle Physics (Snowmass'2021)\\
  \vskip 0.1in
  \rule{\hsize}{0.01in}\\
  \rule[+0.2in]{\hsize}{0.01in}\\[-2em]
\end{center}
}

\usepackage[firstpage=true]{background}
\backgroundsetup{contents={\parbox{6.5in}{\snowmass\editors}},
scale=1,placement=top,opacity=1,color=black,position={3.25in,1.2in}}

\usepackage{fancyhdr}
\fancypagestyle{plain}{%
  \fancyhf{}%
  \fancyhead[C]{}
  \fancyfoot[C]{\thepage}
}

\fancypagestyle{empty}{%
  \fancyhf{}%
  \fancyhead[C]{{\it Snowmass'2021 AF02 Accelerators for Neutrinos - Cyclotron Workshop}}
  \fancyfoot[C]{\thepage}
}
\title{Report of the Snowmass’21 Workshop on High-Power Cyclotrons and FFAs}

\date{}

\usepackage{authblk}

\author[1,2]{A. Adelmann \orcidlink{0000-0002-7230-7007}}
\author[3,4]{J.R.~Alonso}
\author[5]{L.~Calabretta}
\author[6]{H.~Okuno}
\author[7]{T.~Planche}
\author[1]{M.~Haj~Tahar \orcidlink{0000-0003-2995-2868}}
\author[3,8]{D.~Winklehner \orcidlink{0000-0002-0715-6310}}
 
\affil[1]{Paul Scherrer Institut, Villigen, Switzerland}
\affil[2]{ETHZ, Z\"urich, Switzerland}
\affil[3]{Massachusetts Institute of Technology, Cambridge, MA, USA}
\affil[4]{Lawrence Berkeley National Laboratory, Berkeley, CA, USA}
\affil[5]{INFN - Laboratori Nazionali de Legnaro, Legnaro, Italy}
\affil[6]{RIKEN, Japan}
\affil[7]{TRIUMF, Vancouver, Canada}
\affil[8]{\href{mailto:winklehn@mit.edu}{winklehn@mit.edu}}

\begin{document}

\maketitle

\begin{abstract}

We report the state of the field of ``High-Power Cyclotrons and 
FFAs'' (Fixed Field alternating-gradient Accelerators) as discussed by international 
experts during a three-day workshop of the same name.
The workshop was held online Sep 7 to Sep 9, 2021 with 50 registered participants,
as part of the US Snowmass’2021 community exercise;
specifically, the Accelerator Frontier (AF) and the subpanel Accelerators 
for Neutrinos (AF02).

\vspace{5pt}
\noindent{\bf Workshop Charge:}
To take stock of the world inventory of high-power cyclotrons and FFAs, 
to asses available beam currents and beam powers, 
and to investigate limitations. 
Furthermore, to evaluate the role of cyclotrons in particle physics, 
directly used or as injectors to other machines. 
Finally, to discuss novel concepts to push the power,
and provide recommendations to the particle physics and accelerator physics communities.

\vspace{5pt}
\noindent{\bf Findings:}
Cyclotrons (accelerating hadrons)
have played a major role both in nuclear physics and in particle physics
ever since their invention by E.O. Lawrence in 1930. The relativistic increase of 
inertia limits their maximum energy to $\mathcal{O}$(1)~GeV, thus other accelerator 
types have supplanted them at the \emph{energy frontier}. 
However, due to their ability to
provide cw beams of high current, they are very relevant at the \emph{intensity frontier} -- producing copious amounts of pions, muons, and neutrinos at higher energies and neutrinos from 
isotope decay-at-rest at lower energies. 
For example, the PSI proton facility can deliver up to 
2.4~mA at 590~MeV (a 1.4 MW beam), enabling a vibrant muon program.
IsoDAR is designed to produce 10~mA at 60~MeV (a 600~kW beam), producing neutrinos 
at a rate equivalent to 50 kilocuries.
In addition, cyclotrons have \emph{high societal benefit}
through medical isotope production and energy research.

We found that there have been several breakthroughs in the past years to further 
increase the available beam currents (and thus total delivered power) that make
continuous wave (cw) isochronous cyclotrons the accelerator of choice for many 
high power applications
at energies up to 1~GeV. Key innovations are: Improved injection (through RFQ direct 
injection, transverse gradient inflectors, and magnetic inflectors), improved 
acceleration (utilizing \emph{vortex motion}, single-stage high energy designs,
vertical excursion FFAs), and improved extraction
(through new stripping schemes and by \emph{self-extracting}, using 
built-in magnetic channels). The use of \htp as accelerated ion instead of 
protons or H$^-$ has also received much attention lately. Here, stripping
the electron during extraction or directly after doubles the electrical beam current 
mitigating some of the space charge issues with high current beams in the accelerator.

There are now several projects designing new powerful cyclotrons for particle
physics, medicine, and accelerator driven systems (ADS) for energy research.
These are cost-effective devices with small facility footprint, thus following 
the mantra \emph{better, smaller, cheaper}. Among them, the IsoDAR compact cyclotron
promises a 10~mA cw proton beam at 60~MeV/amu, improving by x4 over PSI injector 2 and
by x10 over commercial cyclotrons for isotope production. A design for a 2~mA 
superconducting cyclotron is underway at TRIUMF, further reducing the footprint.
Several designs (AIMA, DAE$\delta$ALUS, TAMU) are being developed for ADS and
particle phyiscs (CP-violation in the neutrino sector).

Finally, we found that the field of computational (accelerator) physics has made great
strides and high fidelity simulations have become a necessity to understand and design
accelerators with high space charge. High performance- and exascale computing
will be needed in order to accurately simulate many-particle interactions (e.g., space-charge 
and halo-formation), and beam-environment interactions (e.g., residual gas, wakefields).
As in other fields, Machine Learning can play a big role by providing new 
tools to understand and predict complex behavior, and significantly reduce
simulation execution time, enabling virtual particle accelerators and 
faster and better optimization.

\vspace{5pt}
\noindent{\bf Recommendations:} We, the community of particle physicists, particle
accelerator physicists, and funding agencies, should:
\begin{enumerate}[topsep=2pt,itemsep=2pt,parsep=2pt]
\item Recognize the important role cyclotrons are playing in Nuclear- and Particle Physics;
\item Encourage development of this type of accelerator, as an investment with high 
      potential benefits for Particle Physics, as well as outstanding societal value; 
\item Recognize and encourage the high benefit of collaboration with the cyclotron 
      industry.
\item Recognize the opportunities the Exascale computing era will provide and adjust
      development of beam dynamics simulation tools accordingly.
\item Aim for a close connection of traditional beam dynamics models with 
      (1) machine learning (surrogate models) and 
      (2) feedback (measurements) from the accelerator, as they
      will pave the way to an intelligent accelerator control and on-line optimisation framework.
\end{enumerate}
\end{abstract}

\clearpage

\setcounter{secnumdepth}{4}
\setcounter{tocdepth}{4}
\tableofcontents

\clearpage

\chapter{Introduction}

\chapterauthor[1]{J. Alonso}
\chapterauthor[1]{D. Winklehner}
\\
\begin{affils}
  \chapteraffil[1]{Editor}
\end{affils}


Ever since its invention by E.O. Lawrence in the early 1930's, the cyclotron has been a workhorse for ``high energy'' physics. For many years the definition of ``high energy'' being pegged to the highest energy available from the particle accelerators of the day.  
The early cyclotrons were able to overcome the Coulomb barrier and thereby give birth to the field of nuclear physics.  After WW II, the 800~MeV 184-Inch Synchrocyclotron at the ``Rad Lab" in Berkeley had sufficient energy to create pions, so became the first machine to contribute to ``particle physics.''  Though the 184" has long-since been decommissioned, and the ``energy frontier" is no longer populated with these machines, cyclotrons of roughly the same energy at PSI, TRIUMF and St. Petersburg have continued to be productive sources of pions and muons.   
In the last few years, new applications for cyclotrons are emerging that can reinvigorate their mission in particle physics: as sources of neutrinos.  Two experiments, described in section~\ref{section:particle-physics}, highlight cyclotrons in this role.


\section{Workshop Charge}

The initial Charge for the Workshop was 
``to summarize the state-of-the-art in high-power cyclotrons and FFAs and 
discuss future (needed) developments,” particularly as high-current 
injectors for other accelerators.
This charge was expanded to cover the current applications of these cyclotrons, 
both in particle and nuclear physics, and in areas of societal relevance, particularly 
medical applications and accelerator-driven systems (ADS) for energy production and 
waste-transmutation. 
The latest innovations in accelerator physics and cyclotron
technology reported here show that an increase in beam current up to a factor 
10 is possible, giving birth to a new generation of compact machines that can be 
applied in many fields (as, for example, evidenced by the IsoDAR cyclotron for 
neutrino physics, or the Innovatron for medical isotope production).

\section{Workshop Program and Participants}

\subsection{Program}

The Workshop took place via zoom on September 7, 8 and 9, 2021. This section will 
provide an overview of the Workshop, the following sections will summarize the
presentations, discussions, and conclusions of each session.

The first day focused on the current state-of-the-art with a description of existing
facilities, and an assessment of the limits of their performance. These sessions were
convened by Luciano Calabretta (INFN-Catania) and Thomas Planche (TRIUMF).
The second day addressed applications of high-power cyclotrons in medicine and 
particle physics, and their role in ADS applications.  Sessions were organized by 
Jose Alonso (MIT), Daniel Winklehner (MIT), and Malek Haj Tahar (PSI). The third day
addressed novel concepts for high-power cyclotrons, including developments in FFAs, 
a new integrated commercial cyclotron package being developed by IBA, efficient
injection developments for compact cyclotrons, and advances in simulations and
machine-learning; this session was organized by Hiroki Okuno (RIKEN) and 
Andreas Adelmann (PSI). The program of talks is listed on the Indico website,
\url{https://indico.mit.edu/e/cyclotrons},
and slides are available by navigating to ``Timetable'' then ``Detailed view.''

\subsection{Participants}

A total of 50 registrants from 23 universities and laboratories, and one company participated in the Workshop.  The list of participants and their affiliations is included at the end of this document.

\section{The Role and Place of the Cyclotron in Particle Physics}
\label{section:particle-physics}
High-intensity cyclotrons and their applications discussed in this workshop fell into two broad energy categories:  higher energies, 500~MeV and above; and lower energies, below 100~MeV.  The higher energy range was discussed mainly in the context of Accelerator-Driven Systems (ADS), for instance driving thorium reactors to criticality for energy production.  
The lower energy cyclotrons are major players today in the production of 
radioisotopes: for medical diagnostics and therapy; or tracers in various 
industrial and research applications.

Cyclotrons in both of these energy ranges can have significant roles in particle physics as well.

\subsection{Higher energy}
E.O. Lawrence’s 900~MeV 184-Inch Synchrocyclotron~\cite{Sewell:184} at the “Rad Lab” in Berkeley, completed in the 1940’s, was the first accelerator with sufficient energy to produce pions.  It provided the foundation for the much larger 6 GeV Bevatron~\cite{Lofgren:Bevatron}, where the field of ``particle physics” truly began.

Two of today’s extremely active cyclotron centers are based around cyclotrons initially 
built as ``meson factories” in the 1970’s.  The PSI 590~MeV separated-sector cyclotron 
continues to hold the record for beam intensities from cyclotrons~\cite{PSI:facility}, 
and the 500~MeV TRIUMF H$^-$ isochronous cyclotron also remains vibrantly
active~\cite{TRIUMF:facility}.  Both have pioneered the use of pions for radiation
therapy~\cite{PSI:pions,TRIUMF:pions}, as well as early studies in pion and muon physics.
Both have branched into many fields, from materials research with $\mu$SR (Muon Spin 
Rotation, Relaxation, or Resonance -- using muon spin to look at structural and
dynamical processed in the bulk of a material on an atomic scale), and spallation 
neutrons at PSI; to rare-isotope production, separation and re-acceleration at TRIUMF.  

The proposed DAE$\delta$ALUS experiment~\cite{aberle:daedalus} calls for an 800 MeV/amu ($\htp$) cyclotron sending 10 mA of protons into a carbon target to produce a copious quantity of pions.  The $\pi-\mu-e$ sequence produces a ``decay-at-rest’’ source of neutrinos which, placed at appropriate distances (1, 5, and 20~km) from a large, hydrogen-containing neutrino detector would provide a sensitive measurement of the CP violating $\delta_{CP}$ parameter using the relatively high cross section of the inverse-$\beta$-decay (IBD) reaction.  Note, in evaluating the L/E ratio -- the usual metric for neutrino oscillation measurements --  the DAE$\delta$ALUS decay-at-rest configuration, though involving shorter distances and lower energies, is quite equivalent to the Fermilab ratio.

\subsection{Lower energy}
Cyclotrons in the energy range of 100 MeV and below are most appropriate for producing radioactive isotopes through nuclear reactions — of protons or ions as primary beams, or neutrons as secondary beams — on specific target materials.  Where the particle-physics applications come into play is in the study of the neutrinos from beta-decay of the isotopes produced. 

Experiments set up close to reactor cores form a large part of ``neutrinos-from-$\beta$-decay’’ sector, however this neutrino source is not pure — several hundred isotopes contribute to the neutrino flux — leading to unexplained phenomena such as the ``5 MeV bump’’~\cite{RENO:2018pwo} that make unambiguous evaluation of  reactor-experiment results difficult.

Pure radioactive sources, on the other hand, have clean, well-understood neutrino spectra, offering advantages for precision experiments~\cite{KATRIN,SAGE}.  For many of these experiments, though, the half life of the decaying isotope, or logistics issues of transportation of kilo- (or mega-) curie sources from production to experiment site~\cite{bellini_sox_2013} can be limiting factors.

Cyclotrons can become crucial players in this arena, producing short-lived isotopes that are continuously replenished by the cyclotron beam.  The IsoDAR experiment, for instance~\cite{bungau:isodar, alonso2021neutrino}, provides a pure, high-endpoint energy $\beta^+$ spectrum associated with the decay of $^8$Li produced by neutron capture of $^7$Li.  The 839 ms isotope is continually replenished by the proton beam (with neutrons produced by the 60 MeV protons striking a (Be + D$_2$O) target, producing what is essentially a very pure 40 kiloCurie source with the strength and lifetime controlled entirely by the beam from the cyclotron.

IsoDAR is primarily a sensitive search for sterile neutrinos~\cite{bungau:isodar}, and a study of $\nuebar-e$ scattering, but the technique, even the $^8$Li isotope, can be used for numerous other particle-physics applications~\cite{alonso2021neutrino}, such as precision mapping of the neutrino spectrum to look for shape deviations and what might be characterized as a ``bump-hunt’’, or further evidence for ``beyond-standard-model’’ (BSM) effects such as axion-like entities, Z’ bosons, or the ephemerous ``X17’’ particle~\cite{x17}. 

Furthermore, the output energy of an IsoDAR-like cyclotron need not be restricted to 
60 MeV/amu. As all the innovations that enable high current (and thus high power) and design challenges are upstream of 1.5 MeV/amu in the cyclotron, other cyclotron designs for
10~mA machines at any energy from 1.5 MeV/amu to 60 MeV/amu (and with additional work
potentially also higher) can be spun out quickly. This \emph{family of cyclotrons} opens up additional opportunities for universities and laboratories to perform interesting particle 
physics at reasonable cost and moderate facility footprint.
\vspace{10pt}

\begin{center} 
In sum, the opportunities offered by the cyclotron for \\
research in particle physics are truly broad.
\end{center}

\section{Overview of the Operating Principles and Types of Cyclotrons}
\label{cycltypes}

The basic principle of the cyclotron is that of ions being
forced on circular orbits by a dipole magnetic field and being accelerated by repeatedly
crossing an oscillating electric field. 
The inward magnetic force balances the centrifugal outward force.

\begin{equation}
F = q v B  =  \frac{\gamma m_0v^2}{R}
\end{equation}

(vector directions understood, all orthogonal in appropriate directions) which leads to:

\begin{equation}\label{isochronicity}
\omega_{RF} = \frac{qB}{\gamma m_0}  = \frac{v}{R}
\end{equation}
		
Thus the revolution frequency of a particle is not dependent on the radius, 
as long as the (qB/$\gamma m_0$) term remains fixed.  

As particles gain energy, the relativistic increase of inertia (here included 
with the gamma-factor), requires modification of either the magnetic field or the 
RF frequency to keep particles synchronous with the accelerating electric field.

Designing the magnet to achieve the same revolution frequency at each radius leads to
the \emph{isochronous cyclotron}.
The importance of isochronicity is that a single RF system, operating at a constant frequency can support acceleration of particles at any radius. This is necessary for the highest intensity of beams, where particles reside stably at all radii and are continuously injected and extracted. Only isochronous cyclotrons can deliver continuous wave (cw) beams.
All high-current cyclotrons are thus isochronous.

Another option exists, referred to as a \emph{synchrocyclotron}, in which the RF frequency is
changed as the particles are accelerated. Synchrocyclotrons are not able to accelerate 
cw beams.
They are useful for applications that require low intensity beams, and where optimization of other parameters provides cost or compactness benefits. An excellent example of this is the 
Mevion 9~T superconducting proton synchrocyclotron that produces 250~MeV protons in a package compact enough (outer diameter of magnet is less than 1.5 meters) to fit on a gantry for delivery of beams for proton therapy.

Vertical focusing is crucial for high-intensity cyclotrons.
A  magnetic field geometry that enables this is the ``Axially Varying Field’’ (AVF), 
where narrow-gap, high-field ``hills’’ are alternated with large-gap, 
very low field regions ``valleys.’’ Valley spaces are used for RF cavities, 
instrumentation, and vacuum pumping systems.
The number of variations around one orbit is often referred to as ``flutter.’’  
Straight radial interfaces with no axial variation provide vertical focusing, 
but spiraling the hills also can provide a component of additional radial focusing.
An extreme magnetic configuration, the ``separated sector cyclotron’’ has each hill section being a separate magnet, with the ``valleys’’ being spaces between the magnets.  
The configuration where the entire machine is surrounded by a single set of coils, 
and the field geometry is determined entirely by iron configuration, is referred 
to as a ``compact cyclotron.’’


Injection of particles can either be done axially, by bringing the beam down the central axis of a compact cyclotron and through a ``spiral inflector’’ that bends the particles into the plane of the cyclotron and directs particles to the first accelerating cavity, or by radial injection where the beam is directed from outside the cyclotron towards the central region and bent into the lowest cyclotron orbit by magnetic or electrostatic deflection.
Many compact cyclotrons also use an internal ion source, often of the Penning (PIG) type
where the compressing magnetic field is provided by the cyclotron itself.

Acceleration is accomplished with RF cavities tuned to the revolution frequency of the particle, or to a harmonic of this frequency. The harmonic number determines the number of bunches around the circumference of the orbit. 

Extraction can either be via stripping foil (for ions whose charge state can be changed by passing through a thin, e.g., carbon foil) or by an extraction channel defined by a thin septum as the inner conductor of an electrostatic deflector.  
For septum extraction, especially for high-current cyclotrons, very good turn separation must be established so a minimum of beam is lost on the septum.  
Such beam loss produces activation, and destructive thermal or erosion damage to the septum.  Good turn separation requires high RF accelerating voltages at the outer regions of the cyclotron, good control of bunch shape, and often the employment of structure resonances.  
Foil extraction is an excellent option, particularly for H$^-$ beams, where every ion entering the foil is converted to a proton that is bent away from the center of the cyclotron.  
Clean turn separation is not needed, and variable energy can be obtained by moving the foil to a different radius, intersecting particles at different stages of acceleration.  
A beam-current limit is determined by the ``convoy electrons’’, the two electrons stripped from the H$^-$ ion that are bent on tight orbits and pass repeatedly through the foil transferring all their energy into heating of the foil. This current limit is around 1~mA.

An important consideration is the vacuum in the cyclotron.  As the distance traveled by each ion is very long, interaction with residual gas is an important source of beam loss.  Another source of beam loss is Lorentz stripping: the relativistic conversion of the rest-frame magnetic field into an electric field seen by the ion.  At high velocities this electric field can be sufficient to strip the electron from an H$^-$ ion, or dissociate the most loosely-bound vibrational states from an H$_2^+$ ion.  These effects limit the upper cyclotron energy to about 70 MeV for H$^-$ ions, or a few 100 MeV for H$_2^+$ ions.

\chapter{State-of-the-Art\label{sec:limitations}}

\chapterauthor[1,2]{L. Calabretta}
\chapterauthor[1,2,3]{T. Planche}
\chapterauthor[1,3]{M. Haj Tahar}
\chapterauthor[1,3]{H. Okuno}
\chapterauthor[3]{C. Baumgarten}
\chapterauthor[3]{J. Grillenberger}
\chapterauthor[3]{J. Kim}
\\
\begin{affils}
  \chapteraffil[1]{Editor}
  \chapteraffil[2]{Convener}
  \chapteraffil[3]{Speaker}
\end{affils}

\section{Introduction}
The classical cyclotron was invented and developed for research in nuclear physics. 
The first major evolution of this type of accelerator was the introduction of the 
azimuthally varying field (AVF) cyclotron, 
otherwise known as the isochronous cyclotron~\cite{heyn_operation_1958}.
Furthermore, the development of computers and superconductivity produced a further 
broad band of cyclotrons of different types tuned for different researches in the 
field of nuclear physics but also for a wide range of applications. 
The golden age of the cyclotron was the period from 1960 to 1990 when many cyclotron 
projects were studied, financed and built. 
Some examples are 
LBNL~\cite{gardner_production_1948, burfening_positive_1949, kelly_general_1962},
DUBNA~\cite{gulbekyan_new_1992, gikal_recent_1998, gikal_ic-100_2008}, 
GANIL~\cite{ferme_status_1981},
MSU~\cite{blosser_michigan_1979, resmini_design_1979},
iThemba Labs~\cite{botha_operation_1989},
RCNP~\cite{ikegami_rcnp_1989},
and RIKEN~\cite{yano_status_1989}, 
just to remember the largest and most famous laboratories. 
They were often equipped with more than one cyclotron, aimed mainly at research in the 
field of nuclear physics and in the projects of synthesis of Super Heavy 
Elements~\cite{dmitriev_she_2020, yano_ri_2001}. 
A special mention goes to the two large cyclotrons of PSI 
(Switzerland)~\cite{PSI:facility} and TRIUMF 
(Canada)~\cite{TRIUMF:facility}, laboratories which delivered the first beam 
in 1974 and 1975 respectively.
These two machines deliver proton beams with a maximum energy of 590~MeV and 520~MeV, respectively, and were built to feed the so-called Meson Factories.
Despite the fact that the initial design beam currents were only 100~$\mu$A 
and 50~$\mu$A, respectively, today they have significantly exceeded their initial target. 
In particular, the PSI machine is able to supply proton beams with currents up to 2.4~mA 
and probably even more in the future.

In this section we host two talks on the status of the art of these two machines. 
The survey of the RIKEN laboratory illustrates the flexibility of 
cyclotrons operated in cascade, up to 4 cyclotrons including the largest superconducting cyclotron. In this scheme, cyclotrons are used to accelerate heavy ions (up to uranium) achieving a maximum energy of 345~MeV/amu to produce radioactive beam well outside of the valley of stability~\cite{yano_ri_2001}.
The talks presented at this workshop show the different critical elements for each complex.
Moreover, a talk presented by Jongwon Kim (IBS, South Korea)~\cite{kim:isol1} 
describes how a new generation of commercial cyclotrons, the 
IBA Cyclone-70~\cite{IBA}, 
can be used to drive an ISOL facility. The latter aims at producing
radioactive isotopes and perform experiments at the extreme limits 
of nuclear physics.

Unfortunately, due to time limitations, we could not hear presentations 
from all of the places where high-intensity cyclotrons are employed or 
developed. Particularly, industry partners
involved in the construction of cyclotrons 
dedicated to proton therapy centers and to the production of medical
radioisotopes were underrepresented
(a list of cyclotron companies can be found in
Appendix~\ref{sec:companies}). 
As a side note, more than 1300 medical cyclotrons are in 
operation worldwide~\cite{noauthor_ptcog_nodate, iaea_cyclotrons}
and proton therapy centers to treat cancer might be considered one of 
the most beautiful applications of cyclotrons nowadays. 
However, these cannot be considered ``high-power''.

This being said, we believe that the following presentations give a 
good overview and, more importantly, let us examine the limitations of 
the current state-of-the-art and how these may potentially be overcome.
Furthermore, we feel that this introduction and references provided here 
and in the appendix paint a fairly complete picture of the world inventory 
of high-power cyclotrons.


\section{Presentations}
\subsection{Talk 1: ``TRIUMF Cyclotron(s) Current Limit''}
Speaker: T. Planche, TRIUMF
\\

According to the experience at TRIUMF, the current limit in compact cyclotrons is a consequence of the small vertical focusing at the
inner radii. Note that TRIUMF 520~MeV cyclotron -- despite its
appearance -- is a compact cyclotron in the sense that the
radius of its injection orbit is comparable with the magnetic gap.
The vertical focusing in the low energy region of such cyclotrons 
is that due to the RF field, which is phase dependent. 
The current limit is reached when the vertical focusing nearly
vanishes, leading to beam losses on the vertical apertures. 
To good approximation, the phase acceptance decreases linearly 
with the peak current. The limit posed by TRIUMF is an accelerated
peak current of 8~mA that means a  0.8~mA average current assuming
an phase acceptance of 36\degree~RF.

\begin{figure}[tb]
    \centering
    \includegraphics[width=0.5\textwidth]
                    {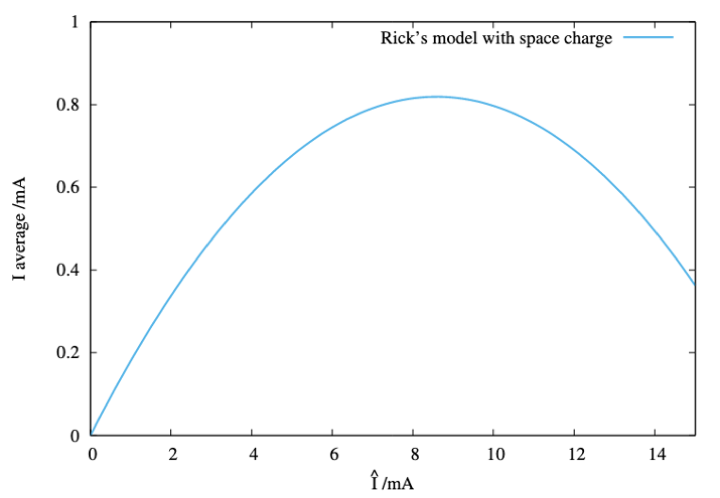}
    \caption{ {\footnotesize According to Ref.~\cite{baartman_intensity_1995}  the average current limit for the TRIUMF accelerator is about 0.8 mA at a Dee voltage of 100 kV. But the 4\degree loss of phase acceptance for each mA of peak current is not known precisely at the moment.} \label{fig:baartman}}
\end{figure}

Commercial compact machines are able to achieve average current on the order of 
1~mA, but they meet additional problems related to the erosion of some components 
of the central region. For these commercial cyclotrons the injection energy is 
usually 35--40~keV.
According to the speaker, increasing the value of the vertical tune 
$\nu_z$ up to 1, could make it possible to accelerate an average beam current 
of 5~mA. This could be achieved quite easily using a separated sector cyclotron 
and/or increasing the injection energy.

\subsection{Talk 2: ``Operational experience with the RIKEN 
            RIBF accelerator complex''}
Speaker: H. Okuno, RIKEN
\\
\begin{figure}[tb]
    \centering
    \includegraphics[width=0.7\textwidth]{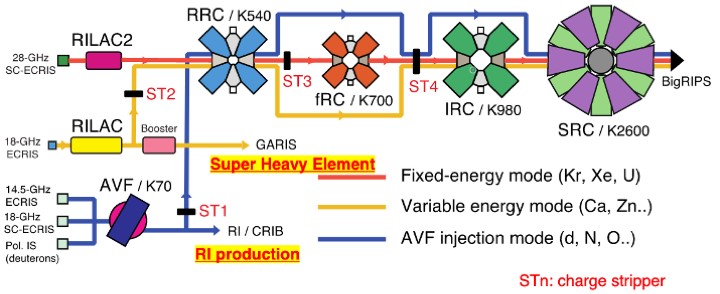}
    \caption{ {\footnotesize Layout of the RIKEN accelerators. From \cite{kamigaito_recent_2020}.} \label{fig:riken}}
\end{figure}

The RIKEN laboratory is the first Laboratory equipped with a Superconducting 
Ring Cyclotron. The laboratory is dedicated mainly to the production of 
Radioactive ion beams accelerating all the kind of ion, from \htp up to $^{238}$U.
The main feature of the RIKEN laboratory is the large number of cyclotrons 
(five cyclotrons in operation) and in particular the ability to drive up to four
cyclotrons working in cascade (see Fig.~\ref{fig:riken})~\cite{kamigaito_recent_2020}. 

\begin{figure}[tb]
    \centering
    \includegraphics[width=0.7\textwidth]{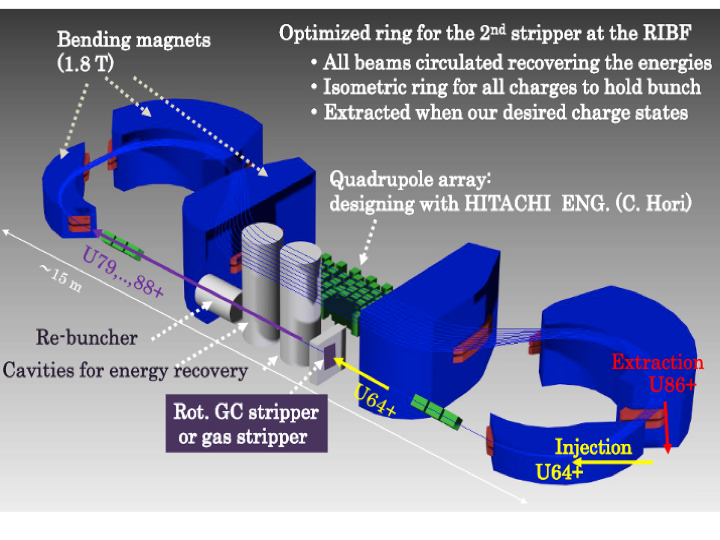}
    \caption{{\footnotesize Layout of the multiple passage Charge Stripping Ring (CSR). 
    From \cite{imao_development_2018}.} \label{fig:csr}}
\end{figure}

It is very difficult to operate the accelerator 
complex when four cyclotrons are connected in series (inject and extract four 
times, energy matching between the cyclotrons, and single turn extraction).
Especially because the four cyclotrons are energy/frequency variable. 
When the ion beam or its energy has to be changed this means retuning all 
4 cyclotrons. Over the years, RIKEN has acquired the experience and know-how 
allowing them to operate their accelerator complex with outstanding reliability.
Operations with ions such as uranium means that a high power-density is deposited
around the surface of the target. This requires developing very careful tuning, 
stable devices, and a fast and reliable machine protection system to prevent 
serious damage to the infrastructure.
Great experience on the use of stripper foils has been learned and the main lesson 
is to use stripper as thin as possible. a new charge stripping concept, the charge stripping ring (CSR), was proposed to perform a multiple passage through the same gas stripper
(see Fig.~\ref{fig:csr})~\cite{imao_development_2018}.

\subsection{Talk 3: ``A 70~MeV cyclotron facility of IBS for ISOL and other uses''}
Speaker: J. Kim, IBS
\\

\begin{figure}[tb]
    \centering
    \includegraphics[width=0.7\textwidth]{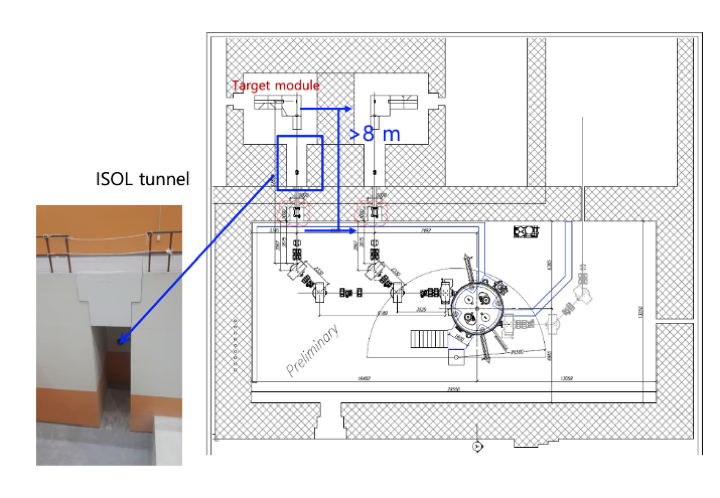}
    \caption{{\footnotesize layout of the ISOL room feed by the beam delivered by the commercial cyclotron Cyclone 70.} \label{fig:ibs}}
\end{figure}

The Institute for Basic Science (IBS) has funded the Rare Isotope Science Project (RISP). This project is based on a commercial compact cyclotron able to deliver proton beams at 70~MeV with intensity up to 750~$\mu$A. The cyclotron was procured and the installation is in progress \cite{kim:isol1}. The main goal is to produce radioactive ion beams with the ISOL method~\cite{jeon:raon1} (see Fig.\ref{fig:ibs}).
A second extraction beam line from the cyclotron will be used by other users for application in nuclear medicine and neutron science communities of Korea.

\subsection{Talk 4: ``The High Intensity Proton Accelerator Facility
            at PSI. Past, Present, Projects.''}
Speaker: Joachim Grillenberger, PSI
\\

The PSI cyclotron complex was the first accelerator to overcome the 
limit of 1~MW accelerated beam. 
It has demonstrated a power output up to 1.4~MW and is now typically operating between 
1.1 and 1.4~MW \cite{PSI:facility}.  
The goal of the original design was to achieve 100~kW. It was quite evident 
from the early years of operations that the ring cyclotron was able to exceed 
this goal.

The main improvement necessary to achieve the 1~MW were:
\begin{itemize}
\item The replacement of the original injector (a compact cyclotron) with a new separated-sector cyclotron;
\item The use of an 870~keV Cockcroft Walton pre-injector to feed the new 
      injector cyclotron;
\item The insertion of a flattop cavity to increase the phase acceptance of 
      the ring cyclotron by a factor four, and reduce the energy spread of 
      the accelerated beam;
\item The replacement of the previous RF cavities of the ring cyclotron
      (made from aluminum) with a new set of copper RF Cavities.
\end{itemize}
The operation at such high beam power and the use of each component near 
their limit poses a set of problems to be taken into account for future 
cyclotrons that plan to operate in the MW territory.
These are the main technical problems recorded at PSI:
\begin{itemize}
\item high dark current at electrostatic deflector
\item exchange required due to damaged insulators
\item insulator coated with thin metallic layer
\item discharges of the electrostatic deflector, but only with RF switched on
\item RF ignites plasma in the proximity of trim coils inside the vacuum chamber induced by the electric field of the RF cavity
\item multipactoring problems in the RF cavities
\item problems with the new flattop RF cavity for the ring cyclotron
\end{itemize}

The new beam dynamics effect (\emph{Vortex Motion}), discovered in the
new PSI Injector 2, allowed avoiding the use of the flattop cavity. 
VM is due to the space charge effect that couples longitudinal and horizontal 
motion and introduces a weak longitudinal focusing.
This effect allows replacing the two RF flattop cavities of the 
PSI Injector 2 with two new accelerating RF cavities.

A problem that is not yet solved is the risk of multipactoring in
the cavities that up today is mitigated by treating the cavities
surface with Aquadag. 

PSI has made serious improvement on many devices as for example on the reliability of the electrostatic deflector, and in general a better reliability can be achieved by a careful planning of repair and service work, continuous replacement of outdated components and “avoiding knowledge drain”.

Other key items are: 
\begin{itemize}
    \item The radiation safety, shielding and waste disposal;
    \item Hot-cell and radio-analytic laboratories;
    \item Licensing and ability to perform required studies;
    \item Formal aspects connected to safety and radiation issues which are often underestimated. 
\end{itemize}

Manage beams with high power exceeding 100~kW pose serious problems
Keys point for a High power cyclotrons are “More diagnostics and simulations” + “well trained operators and more beam development”.
The story of high-power cyclotrons demonstrates that their life is longer than expected and that their applications will evolve along the decade. When design a high power cyclotron infrastructure it is convenient to design the surrounding laboratory with flexibility for future developments.

\subsection{Talk 5: ``Major limitations of fixed field particle accelerators''}
Speaker: M. Haj Tahar, PSI
\\

A brief comparison of cyclical accelerators such as 
cyclotrons, FFAs and synchrocyclotrons in the perspective to
build a high-power machine was presented, showing an improvement of the achieved beam current by three orders of magnitude over the last four decades: The average beam current from isochronous cyclotrons is typically two orders of magnitude higher than synchrocyclotrons, while FFAs, despite their potential for applications beyond the GeV-level, 
have yet to demonstrate their capability for higher currents.

Some of the limitations/challenges on the design of FFAs in general include the accuracy of the dipole magnet constructions (dipole field errors and gradient errors) which can lead to the crossing of resonances \cite{PhysRevAccelBeams.23.054003} and/or imperfect isochronism \cite{Zaremba:1005055}.
Several possible remedies to mitigate these problems were discussed \cite{PhysRevAccelBeams.23.054003,Zaremba:1005055,STAMMBACH19961}. For a given momentum range, the more compact the magnet design is, the more severe the tolerance to imperfections becomes.

Besides, when the aim is to extract a high-power beam, the beam dynamics shall be driven by the need to reduce the losses at extraction to avoid activation problems and allow hands-on maintenance. Several approaches were presented and discussed showing how to achieve this with single turn extraction machines such as the PSI cyclotron.

As a general conclusion, compactness is not necessarily an advantage for high-power machines. Besides, for applications requiring energies at the GeV level or above, FFAs are eagerly awaited to demonstrate their capability to deliver intense beams, at least comparable to what synchrocyclotrons were able to achieve. A summary of the state-of-the-art of FFAs today was presented at the end, as shown in Table \ref{summaryFFA}: synchrocyclotrons, owing to their pulsed mode of operation achieve beam currents that are typically two orders of magnitude lower than that of classical cyclotrons under comparable conditions. For FFAs in general, the situation is less trivial to assess given that the only existing machine is the KURNS scaling FFA \cite{988075} that has a limited current of less than 10 nA for safety reasons related to the ADSR application of which it is the driver \cite{doi:10.1080/18811248.2009.9711620}. 

\begin{table}[htb]
\centering
\caption{\footnotesize State-of-the-art of fixed field light ion accelerators.\label{summaryFFA}}
\begin{tabular}{llll}
\hline\hline
 Concept&Energy reach&Intensity reach&Operation mode\\
\hline
Cyclotron& $\leq 1$ GeV & $\mathcal{O}$(mA) (PSI: 2.4 mA) &CW\\
& at extraction&&\\
\hline
Synchrocyclotron&1 GeV& $\mathcal{O}$($\mu$A) (Dubna: 25 $\mu$A)& Pulsed mode ($\leq$ kHz)\\
\hline
FFA (scaling)& No limit & $\mathcal{O}$(nA) (KURNS: ~10 nA)&Pulsed mode (~kHz)\\
\hline\hline
\end{tabular}
\end{table}

\newpage
\subsection{Talk 6: ``Current Limits of (PSI’s) High Power Cyclotrons: Theory and Practice''}
Speaker: C. Baumgarten, PSI
\\

A lot of interesting and useful simulations have been presented to explain the beam dynamics in the injector and in the ring cyclotron of PSI laboratory. The results of these simulations indicate that it is very important to perform proper simulation to decide the voltage profile vs. radius for the RF Cavities. Also, the effect of strong electric field must be evaluated carefully.
According to the experience with the beam dynamics inside the PSI Injector 2 and inside the ring cyclotron it is mandatory to take advantage of all the information that can be obtained by means of reliable simulations. Nevertheless, there is a reasonable margin of doubt that the reality could be different of the simulations and new phenomena could occur when we go into a higher current regime.
It is important to recall that PSI cyclotrons were built with the goal to achieve 100 kW and that today these machines can deliver power a factor 14 higher. This wonderful result was achieved not only thanks to the engineering margin of the subsystems but also by not usual operations such as switching off the flattop cavities in the injector and off-center beam injection into the ring cyclotron. Fortuitous operations that were well explained afterwards but not investigated at design phase. The construction of a 10~mA cyclotron will be new territory where past experience will certainly be useful but probably will not be enough.
A serious bottleneck pointed by Grillenberger and by Baumgarten is the legal prescription for the beam losses and the problem of activation of components of the accelerator complex and of the vault itself. It was pointed out that the maximum beam current that can be continuously extracted from high power cyclotrons depends on the absolute activation of components. This means that the relative losses shall be lower, the higher the maximum current that is foreseen.
It was also pointed out by Baumgarten that the available formulas for the prediction of the maximum current are too simplified to deliver reliable predictions.

\section{Summary of Cyclotron Limitations}
Looking at the presentations of this chapter we learned that:
\begin{itemize}
    \item Cyclotrons are very long-lived machines: 48, 47, and 36 years old at the PSI, TRIUMF, and RIKEN laboratories,
    respectively;
    \item The development of simulation tools allowed to describe quite well the new beam dynamics phenomena
    (e.g., vortex motion) and the limit of the beam acceptance in the compact cyclotrons anin general the beam dynamics simulation tools are today very safe.
\end{itemize}
The technical limits are mainly due to the original design of these machines.
They could probably be overcome thanks to new technology and to new mechanical design options. The experiences gained along these long years of operation offer us valuable insights to apply to the problem of upgrading.
It is quite evident that the amount of knowledge gained in these years allows starting new projects to achieve higher energy and higher current. The goal of 1~GeV and 10~mA for a proton beam delivered by a cyclotron seems  feasible today, both using conventional technology, or also using superconducting technology to reduce the footprint of the machine (example the superconducting ring cyclotron K2500 of RIKEN).
A problem that must be optimized is related to the reliability and to the maintenance of these new machine. This is a serious problem for accelerator proposed to drive sub critical reactor, so-called ADS. For cyclotrons used in the field of nuclear or particle science this is not a real limit.
As pointed out in the talk by Grillenberger, technical problems like deflector failures 
and their replacement and the reliability of the RF cavities have to be minimized.
For example, some of these problems could be mitigated using robots to replace people in the maintenance operations of the critical components like electrostatic deflectors. Of course, this implies that cyclotron components must be properly designed to allow robot maintenance.
The introduction of robot or automatic maintenance will allow to operate the cyclotron also with larger amount of beam losses due to the higher accelerated current.
Also, the problem to build safer RF cavities avoiding the multipactoring effect need to be optimized. A tentative solution could be replacing the use of Aquadag with alternative coating of the copper. For example, using gold or nickel or other surface treatments. Moreover, careful study of the cavity shapes and using local magnetic field to freeze the multipactoring effect could be an alternative solution to be investigated.
The problem of the limit of the beam acceptance present in the compact cyclotron could be again upgraded using special ferromagnetic materials as Vanadium Permendur that allow to achieve higher magnetic field respect to the classical iron pole and then higher vertical focusing became feasible.
Despite all the technical problems of cyclotrons they are up to now the only “cheaper” solution  to achieve 10 MW proton beam at 1~GeV. 
Indeed, the FFAs discussed in the talk presented by M. Haj Tahar while achieving energies higher than 1 GeV, have not yet a viable solution to achieve the high-power regime. 
Moreover, the machine protection system must be improved not only to protect the infrastructure from serious damage, but also to understand the source of the failure and to allow restarting the accelerator in a short time. Probably this goal can be accomplished using the new tool of machine learning.

The useful information collected by the operations of cyclotrons in the research centers developed worldwide have been received by commercial companies that are today able to supply high-current and reliable machine as the one bought by IBS (Talk 3), to drive their Rare Isotope Science Project. Commercial companies can sell cyclotrons delivering more than 700~$\mu$A of proton beam and new frontiers could be overcome soon (see the talk of G. D’Agostino in the session “Novel concept for high power” of the present workshop).

The scientific community should follow the example of commercial companies to develop and found new cyclotron project at the leading edge. The goal to achieve a proton 
beam with 5~mA at 800-1000~MeV using a cyclotron accelerator is realistic, the critical item to investigate is the best cyclotron configuration to achieve beam currents higher than 10~mA.
According to the recommendations pointed out by C. Baumgarten in his talk, only by pushing toward the limit we will be able to look beyond.

\chapter{Applications}

\chapterauthor[1,2]{J. Alonso}
\chapterauthor[1,2]{D. Winklehner}
\chapterauthor[1,2]{M. Haj Tahar}
\chapterauthor[1]{A. Adelmann}
\chapterauthor[3]{H. Haba}
\chapterauthor[3]{T. Ruth}
\chapterauthor[3]{P. Schaffer}
\chapterauthor[3]{S. Lapi}
\chapterauthor[3]{J. Spitz}
\chapterauthor[3]{F. Meier Aeschbacher}
\\
\begin{affils}
  \chapteraffil[1]{Editor}
  \chapteraffil[2]{Convener}
  \chapteraffil[3]{Speaker}
\end{affils}

Cyclotrons are widely applied today, across many fields where beams of energetic particles are used for inducing nuclear reactions, or for deposition of energy or ionization in different materials.  This section will explore some of the areas where the highest beam currents from cyclotrons are being applied, how today's cyclotrons are addressing the challenges of these applications, and areas where advancing cyclotron performance will be of critical importance to better meet these requirements.

The areas that will be explored are: isotope-production, particle physics, and accelerator-driven systems (ADS).

\section{Isotopes: Medicine to Super-Heavy Elements}

Studying unstable isotopes produced by nuclear reactions is the backbone of nuclear physics research.  Experiments search for new isotopes to explore the limits of nuclear stability, whether drip lines at the edges of proton-rich or neutron-rich extremes, or to the highest-Z region to search for and characterize new super-heavy elements. 

But once the catalog of known isotopes is produced, important applications can be identified for specific isotopes based on their lifetimes, decay characteristics, and chemical properties~\cite{Ruth-overview}.  Most noteworthy applications are in the medical field, where the chemical properties of certain isotopes allow high selectivity for concentration in targeted organs, types of malignancies, or regions of the body, usually  by incorporating the isotope into an appropriate biologically active agents.  Once at the desired site, the characteristic radiations can be used either to image the areas, or to provide a therapeutic dose of radiation to destroy targeted tissue.  

Radioisotopes can be produced either by reactors or accelerators.  
Reactor-produced isotopes~\cite{agency2003manual} come from neutron-induced reactions (absorption or fission) on targets placed in tubes running into or through the core of the reactor.  An important fission example is $^{99}$Mo, the 66-hour parent of the $^{99}$Mo/$^{99m}$Tc) generator, widely used in diagnostic imaging studies.  The target is $^{235}$U (now required to be in a low enrichment form), the $^{99}$Mo is a prominent fission fragment. Most isotopes are produced by neutron capture, (n,$\gamma$) reactions. 
These are all on the neutron-rich right side of the ``valley of stability,'' in the Z vs N Chart of Nuclides, and decay by $\beta^-$ emission. 

\begin{figure}[tb]
    \centering
    \includegraphics[width=0.7\textwidth]{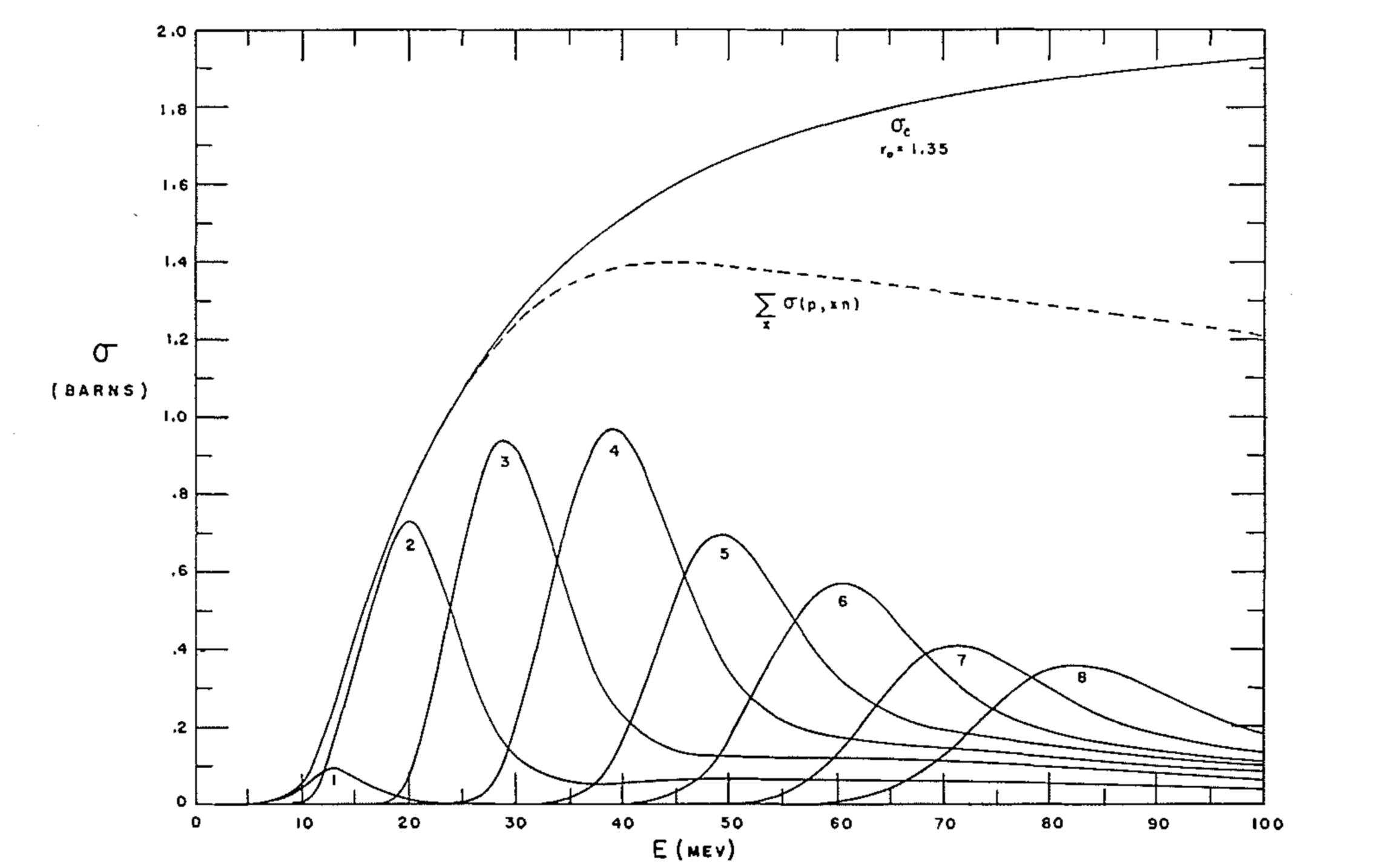}
    \caption{ {\footnotesize Calculated (p,xn) cross sections for $^{209}$Bi bombarded by a proton beam~\cite{JDJackson}.  Protons need about 5 MeV to overcome the ``Coulomb Barrier'' and penetrate through the nuclear surface. Once inside, the various curves indicate the type of reaction that will occur between the proton and the target nucleus. The solid line at the top is the geometric (total reaction) cross section while the asymmetric Gaussian-like curves are the cross sections for evaporation of neutrons, the number of neutrons noted at the peaks. Each neutron carries away about 8 MeV.  Up to about 30 MeV the ``compound nucleus'' (p,xn) process can account for all the reaction products (dashed line), but above 30 MeV other reaction channels, such as the direct knockout of target nucleons or evaporation of charged particles (p or $\alpha$), become important.} \label{pxn}}
\end{figure}

Cyclotrons beams~\cite{schmor:isotopes}, mainly protons, irradiate targets external to the cyclotron, reactions are characterized as A(p,X)B, where target species A is irradiated with protons, resulting in product B and emitting X particles and/or $\gamma$s. 
Using particle beams has the advantage that B is usually a different atomic species than A, so chemical separation of the product from the target can be done yielding ``carrier free'' sources~\cite{carrier-free}.  This allows much higher concentration of the activity in clinical use. It is usually not possible to do this with reactor-produced sources.  

Cyclotron-produced reaction products are almost always on the left side of the Chart, and decay by $\beta^+$ or electron capture.  $\beta^+$ isotopes are of particular value, enabling PET imaging, utilizing the back-to-back 511 keV gammas from the positron annihilation.

Cyclotron beam energy allows selection of specific products. 
Figure~\ref{pxn} shows cross sections or ``excitation functions'' for different reaction products as a function of proton energy~\cite{JDJackson}.  While the calculations are for $^{209}$Bi, this behavior is applicable to almost all nuclei.  
A common experimental technique builds the target up as a stack of thin foils of the same target material~\cite{stacked-foil}, the beam penetrates through a number of layers as it loses energy and finally stops.  Often each foil will show a different activity based on the energy of the proton passing through that foil.

 Most all medically-relevant isotopes are close to the ``valley of stability'' in the Chart of Nuclides, those further away tend to have half lives too short to be of practical interest.  Hence most are accessible with reactions in which 4 or fewer neutrons are emitted from the compound nucleus. Cyclotrons with energies between 10 and 30 MeV are the work-horses of this field~\cite{papash2008commercial}.
 
 A notable exception is $^{225}$Ac, an important therapeutic agent~\cite{Geerlings}.  It can be produced via a (p,2n) reaction on $^{226}$Ra~\cite{apostolidis2005cyclotron}, but the target is extremely challenging:  it is radioactive (1600 year halflife), and only available from reprocessed reactor fuel. An easier production path is using higher energy protons on natural $^{232}$Th, where ``spallation'' reactions produce the desired $^{225}$Ac as one of many reaction products~\cite{robertson2019design}.  The cross section is much lower (around 10 millibarns) than the (p,2n) which peaks at about 200 millibarns, but the target material is much more available, and the desired products can be chemically separated from the target.  Optimal proton energy is about 100 MeV or higher, but the excitation function is very broad.  Efficient production can occur at energies as low as 60 MeV, current production at TRIUMF uses 500 MeV protons.

Medical isotopes are either ``diagnostic'' (for imaging of selected areas of the body) or ``therapeutic''  (for causing radiation damage to localized regions)~\cite{de1998radioisotopes}.  

Diagnostic isotopes are (a) single-photon emitters (100-300 keV), using SPECT  (Single-Photon-Emitting Computed Tomography) imagers)~\cite{bailey2013evidence}, or (b) positron emitters, where the positron stops within a millimeter of the decaying nucleus generating the annihilation photons.  These 511 keV gammas are detected in a ring of photon detectors (called a PET -- Positron Emission Tomography -- scanner) that reconstructs the locus of the decay points, and so the location of highest concentration of the isotope~\cite{basu2011fundamentals}. 
Most often the radioactive atom is 
 attached to a molecule or pharmaceutical that targets a specific site in the body.  For example, FDG or fluorodeoxy- glucose is labeled with $^{18}$F, a positron emitter, and migrates to sites of high metabolic activity. This method is used routinely to identify tumors whether primary or metastatic~\cite{boursi2018functional}.  

Halflife of the diagnostic agent is important. It must undergo minimal decay during transport to the use site, but after the procedure has been completed, usually less than an hour, residual radioactivity only contributes to radiation damage in the patient. Optimally, lifetime of the diagnostic agent should be less than a few hours, making it desirable to have the isotope produced close to the end-use point.  
A common solution is the ``generator"`\cite{lebowitz1974radionuclide} consisting of a long-lived parent whose short-lived daughter is actually used in the procedure. Ubiquitous is the $^{99}$Mo/$^{99m}$Tc set~\cite{arino1975fission}, where the 66-hour $^{99}$Mo parent is transported to the hospital or imaging site, and the 6-hour $^{99m}$Tc daughter is extracted from the generator cell for injection into the patient. The Tc isotope is used for SPECT studies.  A generator of great interest today is the $^{68}$Ge/$^{68}$Ga pair~\cite{rosch2013past}, the $^{68}$Ge parent halflife is 270 days, the daughter $^{68}$Ga is a PET isotope and with a 67 minute halflife.  The longer parent halflife substantially increases the useful lifetime of the generator, and the shorter daughter halflife reduces the radiation dose to the patient. The parent $^{68}$Ge is produced by irradiating natural gallium with protons, via (p,2n) or (p,4n) on the relevant isotope of Ga: 69 or 71~\cite{waites:isotopes}, both about equally abundant. Because of the long parent lifetime, the highest possible proton current must be used to produce sufficiently strong generators.

Therapeutic isotopes are designed to deliver a dose of highly-damaging ionization within a mm or less from where the isotope has been deposited~\cite{lewington1996cancer}.  Either pure $\beta^-$  or  $\alpha$ emitters are extremely effective.  $^{225}$Ac, mentioned above, is a particularly powerful agent~\cite{robertson2019design}.  Being in the heavy ``island'' with uranium and thorium, it decays through many intermediate nuclei to eventually end up in the vicinity of lead.  The decay sequence involves four alpha particles, all with ranges  equivalent to a cell size, and all emitted from the same nucleus.  Place this nucleus in a cell, and you are guaranteed to kill the cell. Long-lived isotopes can be used therapeutically, in a process called ``brachytherapy''~\cite{tanderup2017advancements} where the isotope in a solid container or ``seed'' is  inserted, left  until the prescribed dose is delivered, then removed.  

A recent development has been identification of ``theranostic'' pairs~\cite{zarrintaj2019theranostic}, 
with matched isotopes of similar chemical properties, one diagnostic and one therapeutic.  The patient is imaged first, establishing the selectivity of the carrier to reach the designated target locations, and so the effectiveness of the therapy when the therapeutic isotope replaces the diagnostic one in the carrier.  An example is positron-emitting $^{133}$Ce~\cite{becker2020cross}, serving as the theranostic pair for $^{225}$Ac.  The former is the first of the rare earths, the latter first of the actinides, both share similar chemical properties.  Another theranostic pairs was presented in the third talk of this session:  
two different scandium isotopes, $^{43}$Sc a PET isotope, and $^{47}$Sc a short-range beta emitter~\cite{loveless2021cyclotron}
.

The next section will summarize the requirements and operating characteristics of cyclotrons to be effective in the production of radioisotopes, whether for research or specifically for application in medical procedures.  Following this will be summaries of the three invited papers to this section of the Workshop.

The first will describe the multipurpose cyclotron laboratory at RIKEN in Japan, that supports programs in medical isotopes as well as in a much broader range of nuclear physics applications, including production and identification of new super-heavy elements.  

The second paper describes the very broad isotope research programs at TRIUMF from the perspective of medical applications. 

The third provides a glimpse of a leading university cyclotron center dedicated to the production and distribution of isotopes that are in high demand for present-day clinical applications.

\subsection{Characteristics, Requirements, and Challenges}

Isotope cyclotrons are characterized by high current and reliable operation.  Beam losses above 5 MeV must be as low as possible to prevent activation that hinders maintenance.

As described earlier in Section~\ref{cycltypes}, high current requires that the cyclotron be ``isochronous'', i.e. that the revolution frequency of the ion being accelerated is independent of the radius of its orbit.  This enables a fixed RF frequency to be used, and that particles can be present in the cyclotron at all stages of acceleration.

Isotope cyclotrons mainly accelerate H$^-$ ions, where extraction is accomplished by insertion of a stripper foil, which removes the two electrons leaving a bare proton.  The opposite charge of the proton bends the stripped beam away from the center of the cyclotron, cleanly extracting the beam, with little loss.  The H$^-$ ions are produced in an external ion source and transported through a channel along the central axis of the magnet.  An electrostatic spiral inflector bends the beam into the plane of the cyclotron.

Constraints imposed by using H$^-$ ions are:
\begin{itemize}
  \item Lorentz stripping limits the top energy of the cyclotron~\cite{furman2002lorentz}.  The added electron of the H$^-$ ion is loosely bound (0.7 eV), and the electric field resulting from the relativistic transformation of the cyclotron magnetic field into the rest frame of the ion will become strong enough to remove the electron. For optimally designed compact isotope cyclotrons, the maximum magnetic field is around 2T, and the top energy of the H$^-$ ion should not exceed 70-80 MeV. The highest-energy H$^-$ cyclotron is at TRIUMF, but to reach 500 MeV its maximum magnetic field is 0.5T, and its diameter 16 meters.
\item Stripping foil lifetime limits the maximum current that can be extracted~\cite{waites:isotopes}, also discussed in section~\ref{cycltypes}.  The two electrons removed from the ion spiral inwards in tight orbits less than a mm in diameter, and repeatedly pass through the foil.  For 1 mA of 30 MeV ions, the electrons carry about 32 watts that is all deposited in the foil. Over 90\% of the foil heating comes from these electrons. In most commercial cyclotrons, the total beam current is split between two extraction foils, the maximum of 500 $\mu$A in each foil keeps foil temperatures below the sublimation point of around 4000 K and lifetimes of a few hours.
    \item The vacuum in the machine must be high, to prevent gas stripping of the very large (about 5 times the size of a hydrogen atom) and fragile H$^-$ ion~\cite{milton1995commercial}.  Goal should be at least 10$^{-7}$ torr.
\end{itemize}
 Another factor, not related to H$^-$ but common to all high-current compact cyclotrons, is erosion of the central region~\cite{dehnel2007practical}.  Bunching in the present generation of high-current cyclotrons is very inefficient, only about 10\% of the beam passing through the spiral inflector is captured and accelerated.  The remainder is lost around the central region, causing erosion damage, and facilitating electrical discharges.  Typically the central region of production cyclotrons must be rebuilt about once per year.

Outside of the commercial isotope field, cyclotrons at research laboratories will almost always accelerate beams other than H$^-$~\cite{onishchenko2008cyclotrons}. Most use cyclotrons for accelerating heavy ions, some with cascaded cyclotrons with stripping foils to produce higher charge states after each accelerating stage~\cite{okuno2007superconducting}.   An example is RIKEN outside Tokyo, Japan, whose broadly-based research programs are discussed in the next section.

\newpage
\subsection{Presentations}
\subsubsection{Talk 1: ``Production of radioisotopes for 
               application studies at RIKEN RI Beam Factory''}
Speaker: H. Haba, RIKEN
\\


\begin{figure}[tb]
\centering
\includegraphics[width=5in]{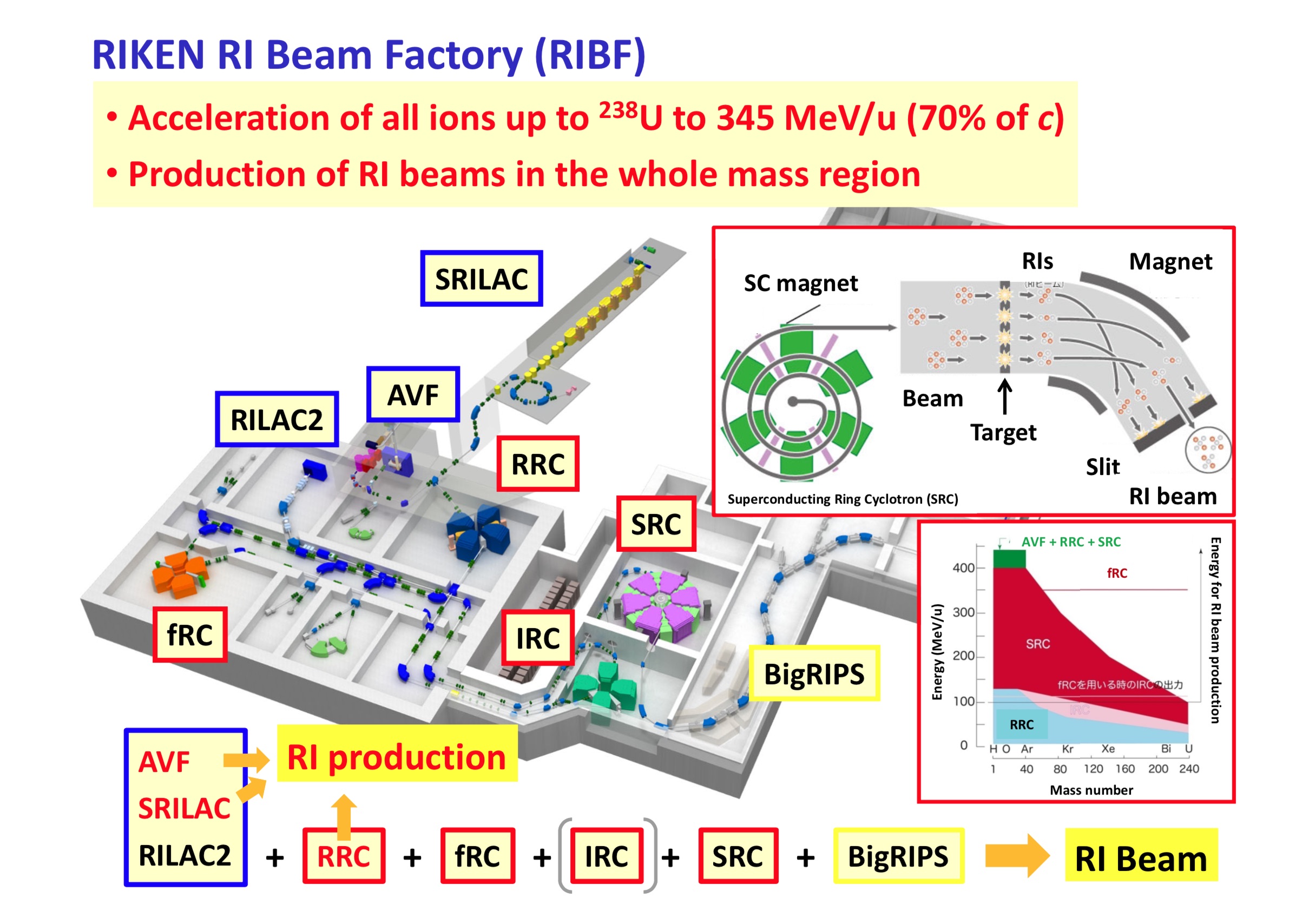}
\caption{\footnotesize Schematic of the cyclotrons at RIKEN's Nishina Center.  From the workhorse AVF K=70 cyclotron to the world's largest superconducting ring cyclotron (SRC, K=2600), the flexibility of this Center's accelerators allows for a large array of programs, from medical isotopes to superheavy elements to radioactive secondary beams.\label{RIKEN}}
\end{figure}

\noindent A schematic of the RIKEN accelerator complex at the Nishina Center for Accelerator-Based Science is shown in Fig.~\ref{RIKEN}~\cite{RIKEN-web}.   With half a dozen independent high charge state (ECR) ion sources, two linacs, and 5 cyclotrons, this Center has unequalled flexibility for generating and tailoring beams of any ion to energies up to 345 MeV/amu.  

The workhorse AVF (4 sector) cyclotron accelerates protons, deuterons, alpha particles and light ions~\cite{watanabe2008calculations}.  Its energy is well suited to nuclear reactions involving compound nucleus formation, and so most of the isotopes used in medical and other reserach fields are generated with this machine.  These isotopes are widely distributed throughout Japan.  In particularly high demand has been $^{211}$At (7.2 hours), used in nuclear medicine~\cite{wang2020present}.

The RRC cyclotron (K-540)~\cite{kase2004present}, produces high current beams of light ions; e.g. ${\mu}A$ beams of $^{14}$N at 135 MeV/amu have been used in ``multitracer’’ studies with a complex targeting system of many thin targets flooded with aerosol-laced helium that collect recoiling nuclei and transfer them to a hot lab area for study, the thin targets followed by a thick target which is processed after irradiation for long-lived products.  Over 50 nuclides of 18 elements are seen with the $^{14}$N beam on natural copper targets.  This technique is very valuable for rapidly characterizing light-ion reaction dynamics and evaluating large numbers of isotopes in a single irradiation.

Superheavy element research is conducted using the RILAC and a sophisticated gas-filled recoil spectrometer (GARIS)~\cite{haba2020present}.  The low pressure gas increases the transmission efficiency of the heaviest atoms, as their orbits are determined by the average charge state of the ion in the gas for their particular velocity~\cite{ghiorso1988recoil}.  Ions recoiling from the target emerge in a wide distribution of charge states, so  are not transmitted well through a vacuum spectrometer.  The ions recoil from the target and are transmitted to a spot on the focal plane of the spectrometer.  The spot can either be instrumented with silicon detectors to observe its stopping point and register subsequent $\alpha$ decays, or can be a small aperture leading to a rapid-chemistry setup for characterizing the individual atoms produced~\cite{haba2007development}. The GARIS spectrometer has been used for chemical studies of Rf (Z=104), Db (Z=105), Sg (Z=106), and Bh (Z=107). In 2004 RIKEN announced the first production of element 113 (Nihonium) from a 5 MeV/amu $^{70}$Zn beam accelerated in the RILAC hitting a $^{209}$Bi target positioned at the entrance of the GARIS spectrometer~\cite{haba2020present}.

An important program at RIKEN is the development of $^{211}$At, a 7.2 hour half-life, $\alpha$ emitting therapeutic agent of great interest in nuclear medicine~\cite{wang2020present}.  It is produced at the AVF using $\alpha$ particle beams of energies below 29 MeV (7.25 MeV/amu) on a $^{209}$Bi target.  At higher energies, $^{210}$At is produced which decays to $^{210}$Po, also an $\alpha$ emitter, but with a 138 day half life.  This long lifetime makes the Po isotope an undesirable background, as it is not easily expelled from the body and contributes an unacceptable added dose to the patient.  The group has been developing high-power targets and in collaboration with many institutions in Japan, radiopharmaceuticals capable of binding the At and transporting it to the targeted tumor site.

RIKEN’s perspective on goals for future cyclotron developments:
\begin{itemize}
\item Cyclotrons which can accelerate various heavy ions ($^{22}$Ne, $^{26}$Mg, $^{48}$Ca, $^{50}$Ti, $^{51}$V, …) with energies of 7 to 10 MeV/amu and intensities around 10 p$\mu$A, for superheavy element research,
\item Low-cost (construction and operation) cyclotrons that can accelerate 30 MeV alpha particles with intensity over 200 $\mu$A for a large-scale, stable supply of $^{211}$At.
\end{itemize}

\subsubsection{Talk 2: ``Radionuclide Production at TRIUMF: The Institute for Advanced Medical Isotopes (IAMI), the Future is Now''}
Speakers: T. Ruth, P. Schaffer, TRIUMF\\

\begin{figure}[t]
\centering
\includegraphics[width=6in]{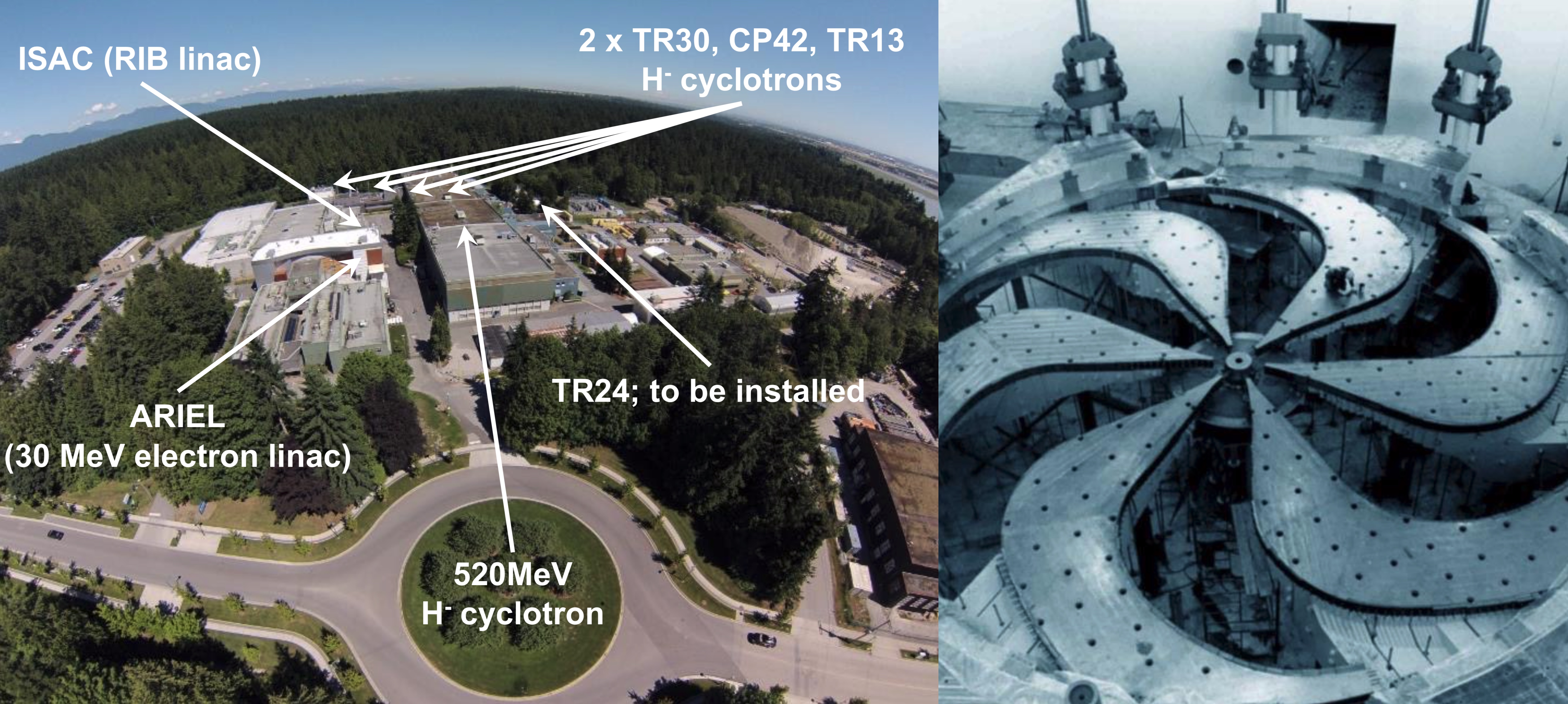}
\caption{ {\footnotesize (Left) Aerial view of the TRIUMF Laboratory showing the various accelerator facilities. (Right) Pole pieces for the 520 MeV H$^-$ cyclotron built in 1968.}
\label{TRIUMFlab}}
\end{figure}

\noindent The TRIUMF Laboratory shown in Fig.~\ref{TRIUMFlab}, Canada’s premier accelerator center has world-leading programs in many fields, particularly beam-based applications in nuclear physics and the life-sciences~\cite{dilling2019fifty}.  Its centerpiece accelerator, the world’s largest H$^-$ cyclotron, delivers up to 350 $\mu$A of energies up to 520 MeV~\cite{bylinskii2014triumf}, was built in 1968 as one three “Meson Factories” (the other two are at PSI and Los Alamos)~\cite{bradamante1988meson}.  It’s life sciences work started with a radiotherapy program based on $\pi^-$ secondary beams~\cite{goodman1990pion}.  

Over the years the inventory of cyclotrons at TRIUMF has grown, there are now five H$^-$ cyclotrons, the four lower-energy machines (13, 2 $\times$ 30, 42 MeV) are dedicated entirely to production of medical isotopes~\cite{hoehr2017medical}.  These four “production” cyclotrons run full-time offering a catalog of over 20 different isotopes for  PET, SPECT diagnostics, or therapeutic agents utilizing short-range $\beta^-$, auger, or $\alpha$ particle radiation.  They are a major supplier of isotopes for North America.  In fact, these cyclotrons were developed by TRIUMF engineers and physicists, and are now manufactured and marketed worldwide by a spinoff company~\cite{acsi_web}.

A major emphasis today is direct cyclotron production of $^{99m}$Tc~\cite{zinkle:scripta}, addressing the shortfall caused by the distribution problems for the reactor-sourced $^{99}$Mo / $^{99m}$Tc generators.  Another is direct production of the PET isotope $^{68}$Ga, providing a second source of this isotope over the difficult-to-obtain $^{68}$Ge / $^{68}$Ga generators~\cite{schaffer2018cyclotron}.

The large 520 MeV cyclotron continues an active career in the medical arena, in both isotope production and proton-beam therapy (including an ocular melanoma treatment facility)~\cite{blackmore1997operation}.  The full energy beam is used either in a dedicated isotope production station, in which a major program now is in producing $^{225}$Ac from $^{232}$Th~\cite{robertson2019design}; or into secondary beam production targets for ISAC~\cite{ball2016triumf} and — just being commissioned — ARIEL~\cite{dilling2014isac}, two sophisticated facilities for isolation and re-acceleration of radioactive species.  These facilities both explore the limits of nuclear stability, as well as characterization of properties of unstable, short-lived isotopes.  Applications are both in nuclear physics as well as life-sciences areas.  ISAC uses only the cyclotron proton beam, while ARIEL can use either this beam or a high-power 30 MeV electron linac as a means of accessing ($\gamma$,X) reactions.

Development of high-power targets has been a major effort~\cite{dombsky2014isac}, as beam power reaches in excess of 100 kW on these stations, and the extremely high radioactivity levels also provides opportunities for development of automated and remote handling capabilities, all directly applicable to radioisotope production.

$^{225}$Ac production via spallation of high-energy protons on natural $^{232}$Th has been a major undertaking at TRIUMF~\cite{robertson2019design}. This Ac isotope, with a 10-day halflife emits four $\alpha$ particles in its decay chain, and is an extremely effective therapeutic agent.  The thick-target spallation reaction at 500 MeV is very efficient, however it also produces a large amount of $^{227}$Ac, which has a 23 year half life, and an almost identical decay chain with four alphas.  Because of its long halflife, $^{227}$Ac is a harmful biproduct, not a clinically useful agent.  TRIUMF has developed a novel technique where, by processing the target with a radium separation procedure a few hours after bombardment, the short-lived $^{227}$Ra will have decayed, while the 15-day $^{225}$Ra isotope then becomes a generator for very pure $^{225}$Ac.  This enables therapeutic doses free of $^{227}$Ac~\cite{robertson2019ac}.  This isotope remains in the waste stream, but is treated as radioactive waste and never reaches the sanitary sewer. (There are severe restrictions on release of $^{227}$Ac to the environment.)

The new Institute for Advanced Medical Isotopes (IAMI) is being completed, including a new TR-24 cyclotron and comprenehsive hot-lab facilities~\cite{bagger2018triumf}.  IAMI  is a multi-story laboratory building dedicated to development of new isotopes and clinically-relevant radiopharmaceuticals, and has collaborative relationships with many local and Canada-wide institutions.

Dr. Ruth’s observations about the IsoDAR cyclotron (10 mA, 60 MeV) and its role in medical isotope production:
It is a unique cyclotron, with unique capabilities.  It is important to look for applications that no other facility could do.  Some examples could be:
\begin{itemize}
\item Producing very long-lived isotopes
\item Exploring new production methods (e.g. gas-phase recoil capture)
\item Production of rare stable/long-lived isotopes.
\end{itemize}

\subsubsection{Talk 3:  ``Isotope Production at UAB (University of Alabama, Birmingham): Expanding the toolbox for nuclear medicine''}
Speaker: S. Lapi, UAB Cyclotron Facility\\

The cyclotron center directed by Prof. Lapi is a mainstream university-based facility dedicated to production and distribution of established radioisotopes, and to development of new radioisotopes and pharmaceuticals for nuclear medicine~\cite{Lapi-web}.

\begin{figure}[tb]
\centering
\includegraphics[width=5in]{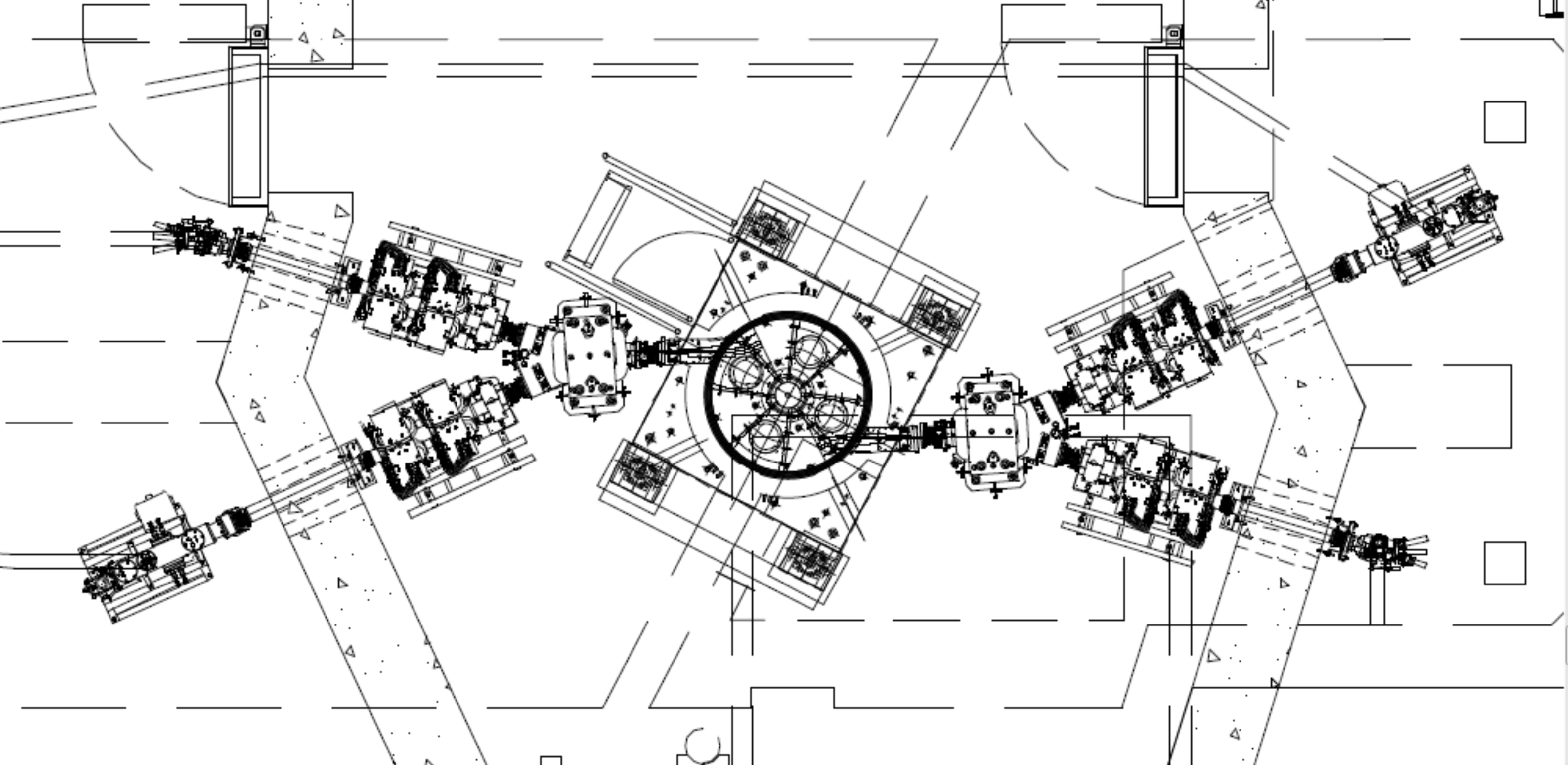}
\caption{ {\footnotesize Plan view of the TR24 cyclotron, its beam lines and target stations.  The pole radius is 60 cm.}
\label{TR24}}
\end{figure}

The center has a single TR-24 cyclotron, produced by ACSI (Advanced Cyclotron Systems, Inc, Vancouver, BC), with extracted beams of 300 $\mu$A (total) at variable energies between 15 and 24 MeV.  The extracted energy is adjusted by careful placement of the stripper foils that convert the H$^-$ to protons, moving the foil to an inner radius for lower energies, but translating it as well so the protons emerge along the proper extraction orbit to reach the target stations.
It uses two foils, 180\degree apart, to produce two extracted proton beams.  Each can be directed to one of two target stations, for a total of four stations, offering different types of targets:  solid, liquid or gas.  Fig.~\ref{TR24} is a plan view of the cyclotron~\cite{watt2015building}.  

The center also hosts PET imaging systems~\cite{mason2021novel}, so short-lived isotopes, $^{18}$F (110 minute), $^{11}$C (20 minute) and even $^{13}$N (10 minute) can be produced and introduced into patients for effective imaging studies.  

Research activities focus on development of $^{89}$Zr~\cite{wooten2013routine}, a 3.3 day PET isotope produced via a (p,n) reaction on a natural (monoisotopic) $^{89}$Y target, and on the theranostic pair $^{43}$Sc (3.9 hour) and $^{47}$Sc (3.3 day), both made by (p,$\alpha$) reactions on titanium isotopes~\cite{loveless2021cyclotron}.  Though (p,$\alpha$) cross sections are not high (maximum of around 25 millibarns), chemical isolation of good quantities of the Sc isotopes is possible, providing very clean samples.  Being the same element, with identical pharmacokinetic profiles, the diagnostic information from $^{43}$Sc will track perfectly the location where the therapeutic $^{47}$Sc will deposit its energy from the short-range $\beta^-$ decay.  This highly-accurate diagnostic verification of a therapeutic procedure is a significant accomplishment.

Prof. Lapi’s Center has cooperative arrangements with many nuclear medicine sites throughout the US and Canada.  Regular shipments of isotopes produced at the Center’s cyclotron are made for both research and clinical programs at the collaborating institutions.


Looking to the future, today's generation of commercial cyclotrons are extremely effective tools for production of important isotopes and offer ample opportunity for R\&D on a host of new isotopes.  In fact, the field has largely been shaped by the capabilities of these machines.  Would a new generation of cyclotrons with higher beam currents offer significant advantages to the field?  The answer is almost surely yes, however at this time it is difficult to see exactly where or how.  Targets capable of withstanding higher beam power will be required, for sure.  But new techniques will undoubtedly be developed that provide advances in the field enabled by such machines.

\subsection{Conclusions}
Cyclotrons have shown remarkable utility in the production of isotopes.  The continuous beam (for the isochronous cyclotrons described), compact footprint, and well-matched beam energy and current to isotope production all contribute to this synergy.  Could there be improvements?  Yes, in several areas.
\begin{itemize}
\item Beam current.  Higher current translates to higher yield, which is always desirable.  Higher current brings with it challenges of more power on the target, increasing needs for better cooling.  In the medical isotope area, cyclotron builders seem to be always ahead of the target engineers; the highest-current present generation of commercial cyclotrons can provide more beam than present targets can handle.  The 1 mA 30-MeV H$^-$ cyclotrons deliver a total of 30 kW, for two simultaneous extraction lines this is 15 kW per target.  This is just at the limit of most of today’s isotope targets.  The new 70 MeV machines double this power.  However, specialized targets are on the drawing boards, and will probably be functioning efficiently in a few years.  This trend gives one confidence that using the power argument against increasing beam current should be viewed as short-sighted.

What are the paths to higher current?  For heavy ions (Z $>$ 1), higher currents come from better ion sources.  This is particularly true for alpha beams, where removing the second electron from a helium ion is very difficult, and accelerating He$^+$ is inefficient.  New  generations of ECR sources are pushing the boundaries of high charge states and currents for lower charge states, so the future is bright in this regard.
For protons (or H$^-$), the limits are in the cyclotron, and understanding beam dynamics in the presence of space charge are the keys to higher current.  As mentioned earlier, H$^-$ cyclotrons are limited by the lifetime of extraction foils, so future advances in Z=1 cyclotron current will most likely come from septum-extracted proton beams.

\item Efficient injection.  Ion sources produce continuous beams, but the phase acceptance of a cyclotron is only about 10\%.  Without compressing the beam, 90\% is lost, pushing the source output requirement to be 10 times higher.  Space charge in the low energy transport line drastically decrease the efficiency of classical RF bunchers.  Developing efficient bunching schemes, for instance using RFQ structures operating at the cyclotron frequency, could be a very great help in increasing the efficiency of ion utilization from the source.  
A further benefit of efficient injection is to prolong the lifetime of the central 
region components and reduce maintenance requirements. Beam lost in the central region 
of the cyclotron causes sputtering and erosion of the copper pieces. Heavily-used 
production cyclotrons may need to have their central region rebuilt every year or so.

\item Reduction of beam loss during acceleration and extraction. Beam loss, particularly for the highest-current Z=1 cyclotrons causes activation and inhibits maintenance access.  Reducing beam losses, in the presence of high currents and significant space-charge forces, requires good understanding of beam dynamics and the mechanisms for beam growth.  Much progress has been made in these areas, as detailed in other sections of this White Paper.
\end{itemize}

\newpage
\section{Particle Physics}
As was discussed in Section \ref{section:particle-physics}, the cyclotron with its
capability to produce high currents of cw proton (and other ion) beams  
with moderate facility footprint has seen renewed interest in particle physics. 

Here we highlight two experiments planning to use a cyclotron driver: the highly
anticipated particle physics experiments IsoDAR and Mu3e. 
In the subsequent discussion section,
we list other ideas for experiments using variations of an IsoDAR-type compact
cyclotron.

\subsection{Presentations}
\subsubsection{Talk 1: ``An application of high power cyclotrons in physics: IsoDAR''}
Speaker: J. Spitz, University of Michigan Physics\\

\begin{figure}[tb!]       
\centering
    \includegraphics[width=0.9\textwidth]{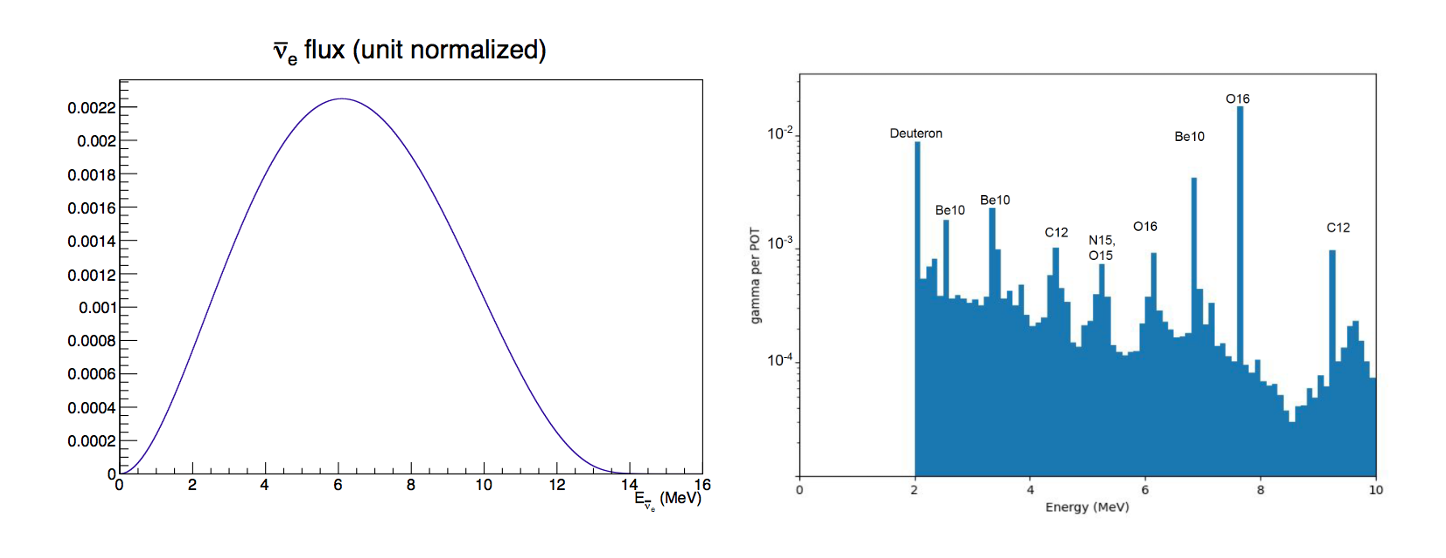}
\caption{{\footnotesize Left: Neutrino flux from the IsoDAR source, unit normalized. Right: photons produced in the IsoDAR target/sleeve, normalized per proton on target.\label{fluxes}}}
\end{figure}

\noindent The original motivation for the Isotope Decay-At-Rest (IsoDAR) experiment 
\cite{bungau:isodar, winklehner2021order} lies in neutrino physics.
Neutrinos exhibit behaviour that was not predicted in the standard model.
They interact with their lepton partners (through the weak force) in their \emph{flavor eigenstate} (\nue, \numu, \nutau), but they travel through 
spacetime in their \emph{mass eigenstates} ($\nu_1$, $\nu_2$, $\nu_3$).
These eigenstates are not aligned and are connected through the PMNS (Pontecorvo, Maki, Nakagawa, Sakata)
matrix.
\begin{equation}
    \begin{pmatrix}\nue\\ \numu\\ \nutau \end{pmatrix}
    =\mathrm{U}_\mathrm{PMNS} \cdot
    \begin{pmatrix}\nu_1\\ \nu_2\\ \nu_3 \end{pmatrix}
\end{equation}
A similar relation holds for anti-neutrinos.
This mixing mechanism leads to neutrino oscillations (the number of 
neutrinos of a certain flavor that are produced at a neutrino source 
might not be the the same that is detected later on) with oscillation
frequencies determined by the squared mass differences of the three mass
eigenstates (e.g. $\Delta m_{12}^2$).
Over the past decades, a number of anomalies have been observed that seem 
to indicate that there may be a new characteristic 
oscillation frequency mode (indicative of a new neutrino state).

The most prominent of these recent results are:
\begin{itemize}[topsep=2pt,itemsep=2pt,parsep=2pt]
    \item {\bf MiniBooNE}, which has observed $\nu_\mu \rightarrow \nu_e$  
          oscillations with a characteristic $\Delta m^2\sim1~\mathrm{eV}^2$
          (4.8$\sigma$)~\cite{miniboone_new} and $\bar{\nu}_\mu 
          \rightarrow \bar{\nu}_e$ oscillations (2.8$\sigma$)~\cite{MB_antinu};
    \item {\bf LSND}, which has seen $\bar{\nu}_\mu \rightarrow \bar{\nu}_e$
          oscillations (3.8$\sigma$)~\cite{lsnd, lsnd2, lsnd3};
    \item {\bf Reactor Experiments} that see electron-flavor
          disappearance anomalies~\cite{reactor}; and 
    \item {\bf Radioactive Source Experiments} that also see electron-flavor 
          disappearance anomalies~\cite{source,2109.11482}.
\end{itemize}

The cross-compatibility of the above results hints at a 
3+1 model, where the three active flavors are mixing with an additional 
``sterile" flavor. However, this is in contradiction with a lack of observed 
muon-flavor disappearance in other experiments (one exception being 
IceCube~\cite{IceCube}). The IsoDAR experiment can decisively address
these open questions with sterile neutrinos, but also yields several
more exciting physics results (see below).

In IsoDAR, a 60~MeV/amu compact cyclotron accelerates 5~mA of \htp
ions that are stripped of the electron to form a 10~mA proton beam 
directly after extraction. The use of \htp during injection and acceleration
alleviates some of the space charge constraints. Additional advancements of
the IsoDAR cyclotron are direct axial injection using an RFQ embedded in the
cyclotron yoke (cf. Sec. \ref{RFQ-DIP}) and utilizing a beam physics effect
in space-charge dominated beams in cyclotrons called \emph{vortex motion}
(cf. Sec. \ref{vortex}).
The resulting proton beam is transported to a target of $^9$Be, producing
neutrons. The neutrons enter a surrounding  isotopically-pure $^7$Li sleeve,
where neutron capture results in $^8$Li. 
This isotope $\beta$ decays with a half-life of 839~milliseconds producing 
a high-intensity decay-at-rest (DAR) $\bar \nu_e$ flux, with peak at 
$\sim 6$~MeV, as seen in Fig.~\ref{fluxes} (left).
Because the source is compact, this can be installed underground next to existing ultra-large hydrogen-based detectors.

\begin{figure}[tb!]       
\begin{center}
{\includegraphics[width=0.85\textwidth]{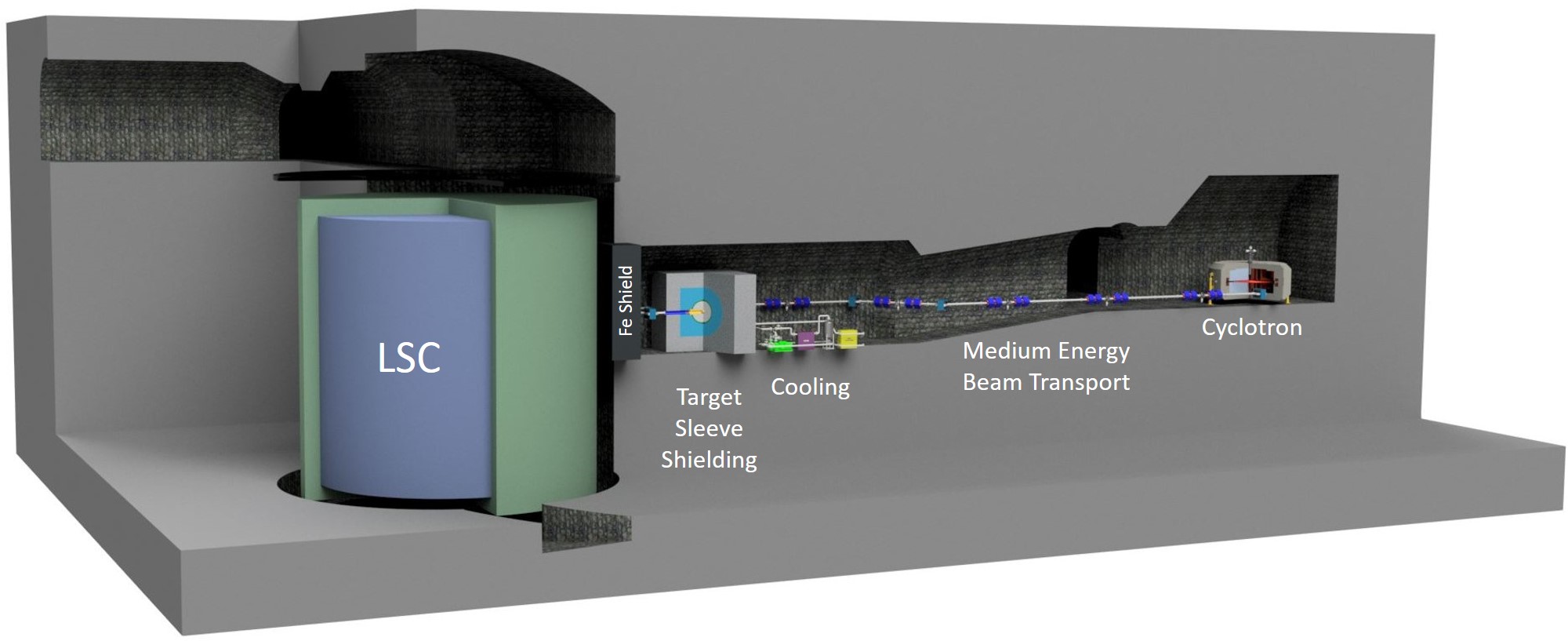}}
\end{center}  
\caption{{\footnotesize Schematic of the IsoDAR neutrino 
         source deployment at Yemilab.\label{yemilab}}}
\end{figure}

\begin{figure}[tb!]       
\begin{center}
\includegraphics[width=0.9\textwidth]{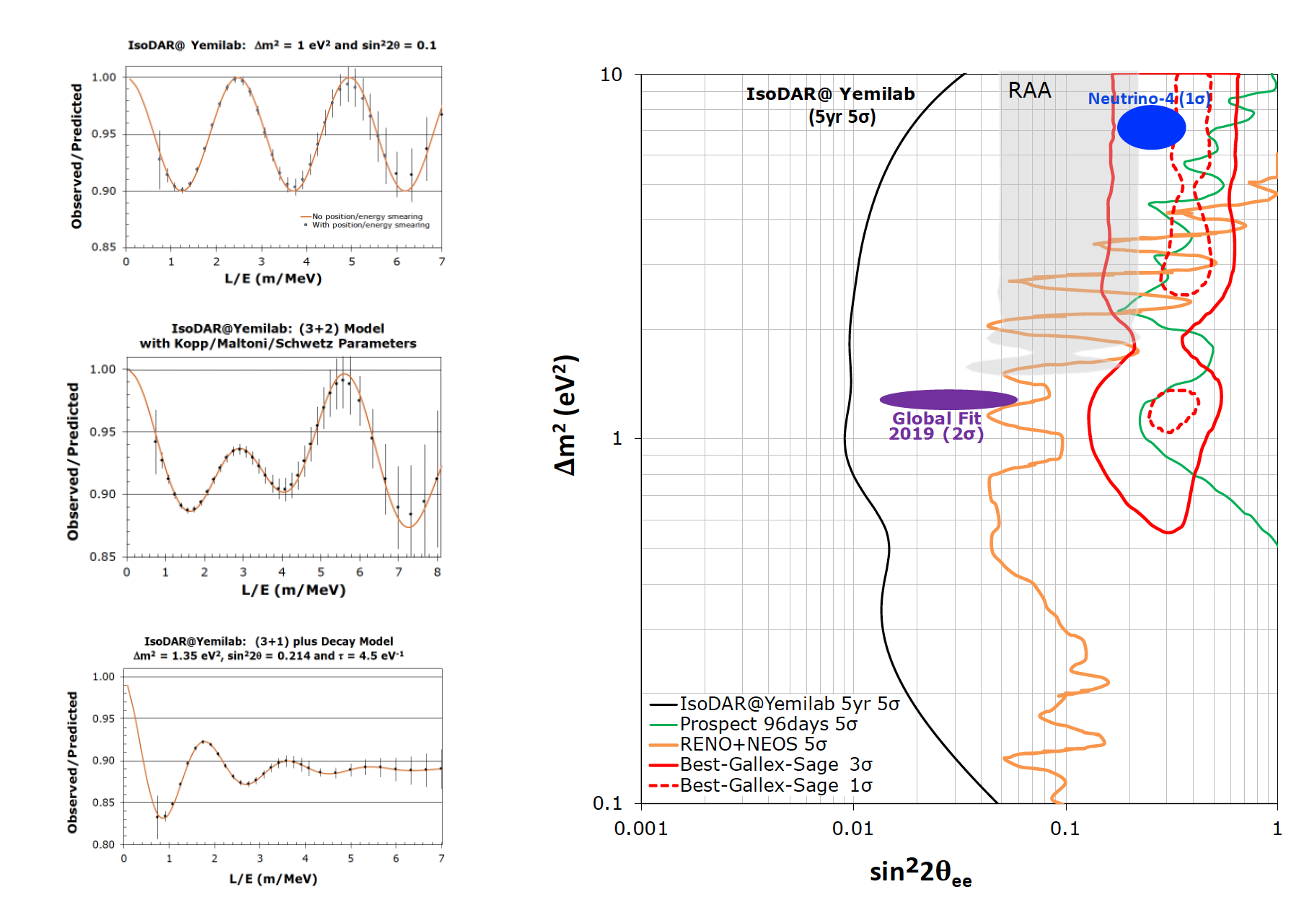}
\end{center}  
\caption{\footnotesize
         Left: Due to the wide range of $L/E$ covered in 
         IsoDAR@Yemilab, many different predictions of $\bar \nu_e$ disappearance
         can be reconstructed. We show three here (a 3+1 model, a 3+2 model, 
         and a 3+1 with decay model). 
         Right: IsoDAR sensitivity in five years of running (four years beam-on time at 
         80\,\% duty factor)
         compared to the global picture of $\nu_e$/$\bar\nu_e$ anomalies (2022). 
         Plots are from Ref.~\cite{spitz:isodar_physics}.
         \label{IsoDARdisappearance}}
\end{figure}

IsoDAR has preliminary approval to run at the Yemilab Center for Underground Physics, 
using the 2.3 kt LSC \cite{seo:yemilab} as its detector (see Fig. \ref{yemilab}).
The main physics cases are presented in detail in 
Ref.~\cite{spitz:isodar_physics} and are three-fold:
\begin{itemize}
    \item {\bf A Search for exotic neutrinos.} IsoDAR's two million inverse beta 
    decay (IBD) events will clear the complex picture regarding the aforementioned 
    anomalies in electron-flavor short baseline oscillations.
    The IsoDAR rates are sufficiently high to accurately trace disappearance signals 
    from $L/E$ of 2 to 10 m/MeV. Fig. \ref{IsoDARdisappearance} (left) shows three 
    examples of signals at Yemilab. Observing any of these would be a smoking gun
    signal worthy to be called ``discovery''. As can be seen in 
    Fig. \ref{IsoDARdisappearance} (right), the sensitivity of IsoDAR is such that,
    within 5 years of running, it covers the most prominent allowed regions and 
    global fits.
    \item {\bf \nuebar-e scattering and non-standard interations (NSI).} 
    IsoDAR's $>7000$ $\bar \nu_e$-$e^-$ elastic scattering (ES) events 
    produced above 3 MeV result in impressive sensitivity to (NSI) couplings. 
    This sample, which is $\times 5$ larger than existing samples, opens new 
    opportunities for purely-electron-flavor precision electroweak studies 
    at low energy, i.e. a measurement of the weak mixing angle $\sin^2{\theta_W}$
    at low Q.
    \item {\bf Bump hunting in the neutrino spectrum.} 
    IsoDAR can also scan the IBD spectrum (rate vs. energy) for peaks that
    are not in the well-understood IBD prediction.
    Such a peak could arise from light-mass mediators (motivated by theory interest),
    a potential new particle called ``X17''~\cite{Krasznahorkay:2015iga} or 
    other BSM effects motivated by anomalies, e.g. a bump at 5~MeV in several
    reactor flux predictions~\cite{prospect2,stereo,NEOS:2016wee,reno,dayabay,doublechooz}.
\end{itemize}

\subsubsection{Talk 2: ``The search for \texorpdfstring{$\mu^+\rightarrow e^+e^-e^+$}{mu -> eee} and what it may need beyond Mu3e phase II''}
Speaker: F. Meier Aeschbacher, PSI\\

\begin{figure}[tb!]       
\begin{center}
\includegraphics[width=0.3\textwidth]{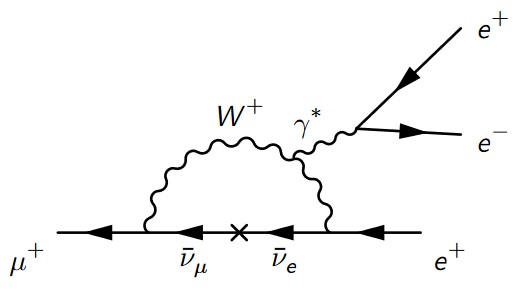}
\includegraphics[width=0.3\textwidth]{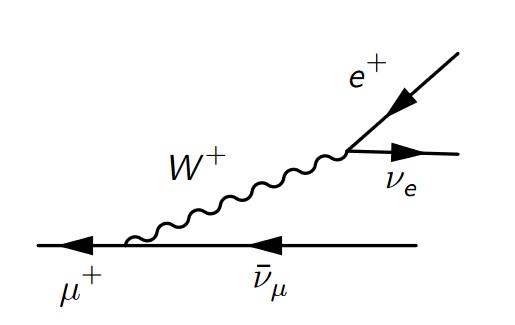}
\includegraphics[width=0.3\textwidth]{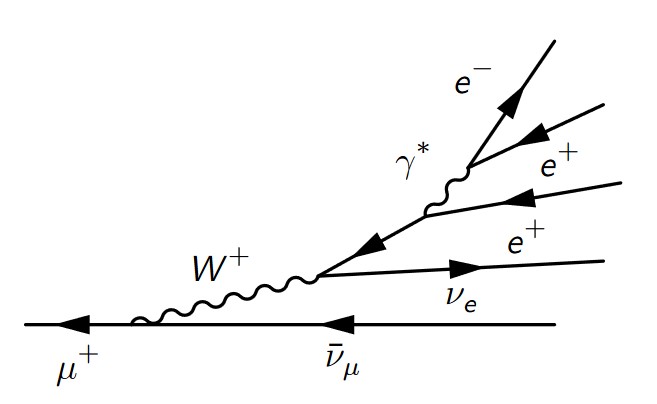}
\end{center}  
\caption{\footnotesize
Feynman diagrams of the desired $\mu^+\rightarrow e^+e^-e^+$
process (left) and the two largest backgrounds. The standard Michel decay (center) has
a standard model predicted branching ratio of 0.99997, and the radiative standard model 
decay (right) has a branching ratio of $(3.4\pm0.4)\cdot10^{-5}$. Courtesy of F. Meier Aeschbacher.
\label{fig:mu3e}}
\end{figure}

\noindent The Mu3e experiment is aimed at observing the lepton flavor changing process
$\mu^+\rightarrow e^+e^-e^+$ 
(see Fig. \ref{fig:mu3e}, left), which has a branching ratio $<1\cdot10^{-54}$ according to 
the standard model. However, alternative models predict a branching ratio of
$<1\cdot10^{-16}$, which would be in the range of an experiment like Mu3e.
The goal is thus to measure $\mu^+\rightarrow e^+e^-e^+$ or exclude
a branching fraction of $>10^{-16}$ at 90\,\% confidence 
level \cite{mu3e_phase1}. The main backgrounds are the standard Michel decay 
(Fig. \ref{fig:mu3e}, middle) and the radiative SM decay (Fig. \ref{fig:mu3e}, right).

Muons impinge on a target in the center of a 1~T solenoid
which forces the resultant electrons and positrons on curved trajectories. Tracks are reconstructed using scintillators and pixel detectors, the common vertex is 
calculated from measurements with the inner pixel layers (vertex tracker).

Assuming a total reconstruction efficiency of 20\,\% in the Mu3e detector, 
these backgrounds have to
be suppressed to the $10^{-16}$ level and more than $10^{17}$ muons have to be stopped in
the detector. 

Muons for Mu3e will be produced at PSI's \emph{Swiss Muon Source}, which is 
driven by the High Intensity Proton Accelerator (HIPA) facility.
Protons at HIPA are accelerated using the 590~MeV separated-sector
cyclotron~\cite{PSI:facility}, fed by Injector II.
\begin{figure}[tb!]       
\begin{center}
\includegraphics[width=1.0\textwidth]{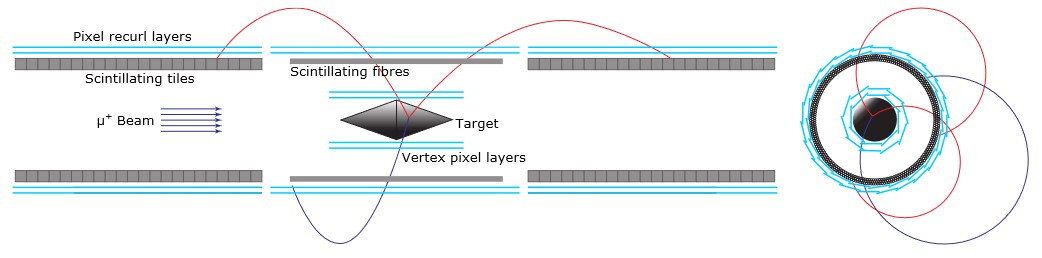}
\end{center}  
\caption{\label{fig:mu3e_det} 
         \footnotesize Schematic of the Mu3e detector layout.
         Left: Side view, 
         Right: Front view. 
         A 1 T solenoid field forces the resultant electrons and positrons produced in the target (center) on curved trajectories. Tracks are reconstructed using scintillators and pixel detectors, the common vertex is 
         calculated from measurements with the inner pixel layers (vertex tracker). Courtesy of F. Meier Aeschbacher.}
\end{figure}

Mu3e is envisioned in three phases:
\begin{enumerate}[label=\Roman*.]
    \item Using the $\pi$E5 beamline, which provides muons stops on target (s.o.t.)
          at a rate of $10^8$~Hz, Mu3e will be able to 
          Installation and commissioning will be performed in 2023 with data taking
          until 2026.
    \item The upcoming High Intensity Muon Beamline (HIMB) at PSI will be able to provide 
          muon s.o.t at $10^9$~Hz, boosting the sensitivity of Mu3e. In order to 
          use this higher rate, Mu3e's pixels and scintillator fibres will have to be 
          upgraded for higher rate. This is ongoing research.
    \item Beyond Phase II, the transport of muons from HIMB to the target 
          inside the detector will be the limiting factor. HIMB is envisioned to 
          ultimately deliver 
          a $10^{10}$~Hz rate, but only a tenth of the muons reach the target.
          In addition, significant further upgrades of the pixels will be necessary.
\end{enumerate}
In summary, Mu3e is an excellent example of highly interesting particle physics -- 
a beyond-standard model search that is driven by a high-power cyclotron. It underlines
why cyclotrons are highly relevant and upgrading the available beam currents beyond
2.4~mA of protons will enable further exciting studies.

\subsection{Discussion}
We do not limit ourselves to the highest available energies at high proton beam currents.
For one, the acceleration of high currents of other ion species will lead to
interesting applications in isotope production.
On the other hand, lower energies (and the associated smaller facility footprint)
will be of further interest for particle physics. Here we present several new ideas for 
such experiments using IsoDAR-like cyclotrons. As the innovative aspects of the IsoDAR
cyclotron are mostly upstream of 1.5~MeV/amu, similar cyclotrons can be designed and built
for any final energy between 1.5~MeV/amu and 60~MeV/amu with minimal additional technical
challenge. For brevity we call them \htct (\htp High Current Cyclotron) here.
\begin{itemize}
\item \underline{Cross Section Measurements with the $^8$Li-based $\bar \nu_e$ flux} -- {\it using an \htct}-60:  As discussed in Ref.~\cite{Tomalak:2020zfh}, new facilities built with an  IsoDAR target/sleeve configuration will permit measurement of the antineutrino cross sections for neutrino coherent scattering for nuclei (CE$\nu$ENS). Multiple detectors of varying materials can surround the target region at a distance of $\lesssim 3$m from the sleeve edge because the target/sleeve design absorbs neutrons with very high efficiency.     The pure electron-flavor flux is relatively sharply peaked at $\sim 6$~MeV (see Fig.~\ref{fluxes}, left) which is complementary to the spallation pion/muon decay at rest sources.

\item \underline{Neutrino fluxes} -- {\it using an \htct}-15:  $\nu_e$ (as versus $\bar \nu_e$) fluxes can be produced 
up to 3.75~MeV by targeting 15~MeV protons on $^{27}$Al to produce  $^{27}$Si.    This was proposed in Ref.~\cite{Shin:2016oxc} for an accelerator-based study of the vacuum-matter transition region relevant for solar neutrino experiments using a radiochemical detector located 10 m from the target.  However, we are most interested in developing this as a potential upgrade for DUNE, since the threshold for $\nu_e + ^{40}$Ar$\rightarrow ^{40}$K$^* + e^-$ is 1.5~MeV.  This would allow measurement of the neutrino cross section at threshold, which is of interest to the cross section community.

\item \underline{Monoenergetic Photons for BSM Searches} -- {\it using an \htct at energy to be optimized}: In designing IsoDAR, we have noted that high rates of monoenergetic photons from decay of excited nuclei are produced and immediately contained within the target.   The photon flux is shown in Fig.~\ref{fluxes}, right.   For a future experiment, the beam energy can be chosen and the target/sleeve material selected for the purpose of producing specific monoenergetic photon peaks of interest use in searches for new physics that couples to photons, such as axion-like and Z$^\prime$ particles. We note that the monoenergetic nature is maintained in conversion to the new particle if the mass is relatively light, so these peaks are valuable for rejecting background.

\item \underline{Detector design, calibration, and testing in high-neutron-flux 
environments} -- {\it using an \htct at energy to be optimized}: 
A local cyclotron facility can allow testing of materials, detectors 
and high power targets. Such a facility would make use of an \htct-60.  
The beam would be extracted as \htp ions, and then partially-occluding 
stripping-foils combined with dipole magnets will be used to break the beam 
into multiple secondary proton beams.
The beam energies can be degraded if energies less than 60~MeV are required.
\end{itemize}

\section{Accelerator Driven Sub-critical Reactors}
The following is a summary of the relevant talks at the workshop that were given by:
\begin{itemize}
    \item F. M\'eot, Megawatt Class Beams From Fixed-Field Rings For ADS-Reactor Application
    \item R. Barlow, ADS prospects and requirements
    \item L. Calabretta, Limits of present cyclotron projects for ADS and future perspectives
\end{itemize}

The Accelerator Driven Sub-critical Reactor (ADSR) is a hybrid system coupling a  particle accelerator to a sub-critical reactor core: the accelerator produces the high power beam which strikes a heavy metal target, in solid or liquid state. The spallation target will thus emit neutrons among other particles in the forward direction of the beam. Such neutrons are fed into the sub-critical reactor core to induce further fission reactions. Owing to the sub-critical state of the reactor, ADSR is considered as inherently safe and shutting down the reactor can be achieved by switching off the high power beam. 

There exits two options that shall render ADSR competitive vis-\`{a}-vis other advanced reactor concepts/fuel cycles:
\begin{itemize}
\item The incineration of nuclear waste in dedicated systems with a large fraction of Minor Actinides (Np, Am, Cm,...): the latter lack the property of producing enough delayed neutrons to control the kinetics of a critical reactor. Hence the need for ADSR.
\item Thorium-fuel cycles where producing the fissile $^{233}$U from Thorium requires the intermediate $^{233}$Pa ($t_{1/2}=27$ days). This induces a time-lagged response in power making the reactor difficult to control in a critical mode.   
\end{itemize}
 
\subsection{Challenges and Requirements}

Despite the potential benefits of ADSR system, several challenges for its development remain such as the accelerator-reactor interface issues and the beam requirements. All challenges are determined by the level of sub-criticality requested for safe operation i.e., by the effective multiplication factor $k_{eff}$. The latter is an average quantity that measures the way the neutrons multiply in the reactor blanket from one generation to the next. In a critical reactor, $k_{eff}$ shall remain close to 1 thus ensuring the steady state character of its operation. However, in an ADSR, $k_{eff}$ is chosen below unity thus enabling more freedom in the type of fuel and its content that can be handled. The lower $k_{eff}$ is, the more dependent the blanket becomes on the external source of neutrons and consequently on the power demanded from the accelerator. Coupling the high power accelerator to an advanced nuclear reactor via the spallation target is a challenging proposition. Some of the difficulties involve the beam current density at the target, its cooling, the shielding requirements especially for the high energy cascade neutrons, the thermal stress induced at the target as well as the structural material survivability. Interested readers are referred to the comprehensive review articles in the literature \cite{rast-2015, AtAbderrahim:2010aa, BAUER2001505}. \\

On the accelerator side, the challenges are driven by the need to reliably achieve and deliver the high beam power at the target, such that the operational aspect of the nuclear reactor facility is not altered. In order to gain better insights into the requested beam parameters, it is instructive to recall the expression relating the thermal power of the reactor core to the beam current $I$ of the accelerator and the multiplicity of the target $N_0$. This writes as follows:
\begin{equation}
    P_{th,c}(MW) = E_f(MeV) I(A) \dfrac{N_0}{\nu} \dfrac{k_{eff}}{1-k_{eff}}
\end{equation}
where $E_f \approx 200$ MeV is the energy released per fission, $\nu \approx 2.5$ is the number of neutrons released per fission process and $N_0 \approx 25$ is the number spallation neutrons per proton assuming a 1 GeV proton beam. Note that the power of the decay heat is neglected in the above expression.\\
Thus, assuming a typical sub-criticality level of 0.95 leads to a requested average proton beam current of 10 mA in order to sustain a 300 MW thermal reactor output. This is one order of magnitude higher than the maximum achieved nowadays, $\approx 1.4$ MW at PSI main ring cyclotron. Taking into account the power drained by the accelerator consumption, $P_A$, the above expression can be further simplified to
\begin{equation}
    P_A \approx \dfrac{1-k_{eff}}{2 \eta_A} P_{th,c}
\end{equation}
where $\eta_A$ is the wall-plug conversion efficiency of the accelerator. Thus, the power drained by the accelerator consumption which lowers the overall efficiency of ADSR is a major challenges for such a concept to compete with other advanced reactor fuel cycle options~\cite{HajTahar:2016}. In this regard, cyclotrons offer the best conversion efficiencies today. \\
Another major challenges for ADSR is the beam reliability requirement which is dictated by the design choices of the core. This has a major impact on the economics/capacity factor of the system and shall be carefully addressed: Primary beam losses from the accelerator taking longer than few seconds cause a drop down of the reactor power to decay heat levels and may induce thermal stress issues. Given that the procedure to recover the nominal power of the core takes several hours, it is vital for ADSR to limit the beam trips as much as possible.

Several ADSR projects are on-going/planned today:
\begin{itemize}
    \item MYRRHA (Multi-purpose hYbrid Research Reactor for High-tech Applications) project at SCK-CEN is the leading project for ADSR in Europe \cite{ABDERRAHIM2001487, de2021belgian}: the aim is to deliver 4 mA average proton beam current from a linac at a final energy of 600 MeV, thus 2.4 MW average beam power to drive a 50-100 MW thermal reactor.
    \item ADANES (Accelerator Driven Advanced Nuclear Energy System) project is planned by the Chinese Academy of Science at Huizhou in China \cite{He:IPAC2019-TUYPLS2}: relying on a linear accelerator design, the objective is to deliver up to 15 mA average proton beam current at 1 GeV and drive a 1000 MW thermal reactor aiming at closing the fuel cycle. 
    \item ADSR project under development at BARC in India: relying on a staged approach, the final goal of the accelerator is to deliver a 1 GeV, high intensity CW proton linac. The latter is being pursued in collaboration with Fermilab. The development of ADSR in India is strongly related to the Thorium utilization program.  
\end{itemize}

The driver of those projects are linear accelerators, despite the major advantage that cyclotrons bring such as achieving the highest conversion efficiencies, having the smallest footprint, and reducing the construction and operation costs. This is mainly due to two fundamental reasons:
\begin{itemize}
    \item The absence of existing cyclotrons with energy in the territory of 1 GeV capable of producing several milliamps of beam current.
    \item The reliability requirement which linear accelerators offered to resolve by means of modular and fault-tolerant designs.
\end{itemize}
In what follows, we discuss some of the potential candidates that shall render cyclotrons/FFAs viable options for ADSR drivers.

\subsection{``Solutions''}
\begin{figure}[t!]
\begin{center}
\begin{minipage}{20pc}
\includegraphics[width=20pc]{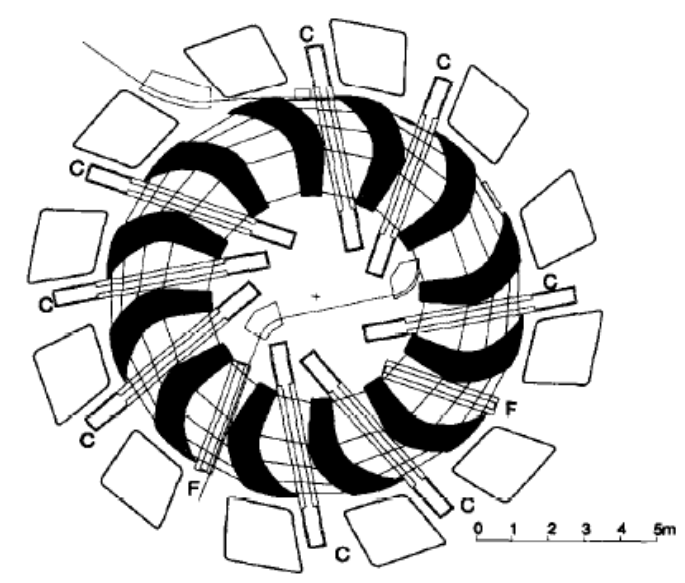}
\caption{\label{PSI-dream-machine} Layout of the so-called ``PSI dream machine'' proposed by W. Joho, PSI. }
\end{minipage}\hspace{20pc}%
\end{center}
\end{figure}

\begin{figure}[t!]
\begin{center}
\begin{minipage}{24pc}
\includegraphics[width=24pc]{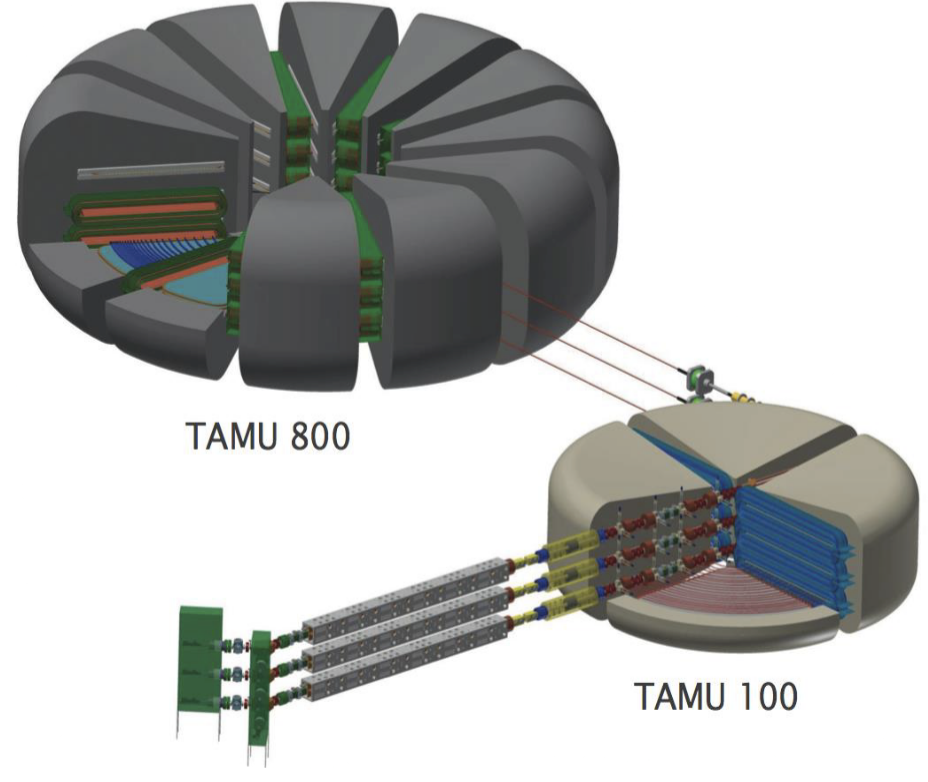}
\caption{\label{TAMU} Two-stage stacked cyclotrons proposed at Texas A\&M University where RFQ accelerators are employed to inject the beam into the smaller of the two ring, the TAMU 100.}
\end{minipage}\hspace{20pc}%
\end{center}
\end{figure}

\begin{figure}[t!]
\centering
\includegraphics[width=0.7\textwidth]{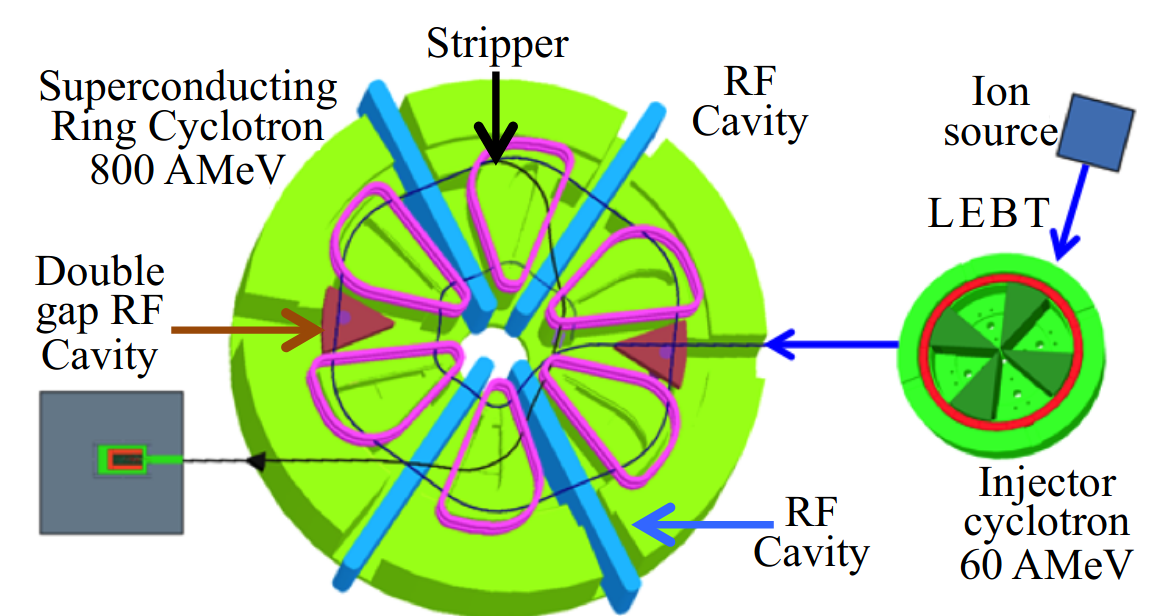}
\caption{\footnotesize Layout of the DAE$\delta$ALUS two stage cyclotron complex.\label{DAEDALUS-ADS}}
\end{figure}

{\label{3.3.2}}
Despite the aforementioned arguments, the perspective for the future is encouraging with several novel concepts based on cyclic accelerators (Cyclotrons/FFAs) that emerged in recent years and that we aim to summarize here. The following cannot give a complete overview of the field and we refer the interested readers to the more detailed discussions on this vast topic \cite{rast_luciano, Mandrillon:Cyclotrons2016-FRA01}.\\
One of the oldest concepts is the so-called "PSI dream machine" that was proposed by T. Stammbach in 1996 \cite{stammbach1996feasibility} and shown in fig. \ref{PSI-dream-machine}. This is essentially a 12-sector scaled-up version of the PSI 590 MeV ring, and aims to deliver 10 mA average proton beam current at 1 GeV. The large number of RF cavities (8) is crucial to reduce the space charge effects and allow enough turn separation for clean extraction by means of an electrostatic deflector. One of the critical aspects of this clean single turn extraction scheme is the reliability of the machine in case one or more RF cavities fail. Operating the machine below the maximum subsystems limits is mandatory to alleviate such a problem.\\
In an effort to overcome the reliability challenge for ADSR, the concept of ``stacked cyclotrons'' was proposed at Texas A\&M University \cite{Assadi:2012zzb}: Aiming to achieve a 10 MW proton beam at 800 MeV, the solution consists of a set of 3 (or 4) stacked cyclotrons and an individual beam current of 4.3 mA (or 3.2 mA) for each. A set of three or four RFQs are used to injects 2.5 MeV proton beam into the first stage of cyclotrons that subsequently accelerate the beam and feed it into the stacked cyclotrons at 100 MeV. This is illustrated in fig \ref{TAMU}. One of the main features of this proposal is the use of superconducting cavities to rapidly and efficiently accelerate the beam. The stacked solution reduces the problem of space charge effects by calling for a more complicated design. However, care must be paid to the reliability issue in case one or more RF cavities fail. In particular, it is not clear whether the RF cavities of one plane are independent of the other ones in presence of perturbations or failures.
 \\ \\
  Using the stripping extraction method, it is no longer relevant to obtain a single
acceleration trajectory, nor well separated turns at the extraction as is the case for proton cyclotrons. In the context of ADSR, two cyclotron proposals aiming to accelerate the molecular hydrogen beam H$_2^+$ stand out: the DAE$\delta$ALUS cyclotron\cite{Minervini2012EngineeringSO} and the AIMA cyclotron \cite{Mandrillon:Cyclotrons2016-FRA01}: the 800 MeV/amu Superconducting Ring Cyclotron proposed for the DAE$\delta$ALUS experiment and consisting of two coupled cyclotrons as illustrated in fig \ref{DAEDALUS-ADS} could deliver a beam power exceeding 10~MW: in the first stage, the beam is accelerated up to  60 MeV/amu and extracted by means of an electrostatic deflector. The second stage consists of an 800 MeV/amu machine capable of producing up to 12 mA proton beam current.

\begin{figure}[t!]
\begin{center}
\begin{minipage}{20pc}
\includegraphics[width=20pc]{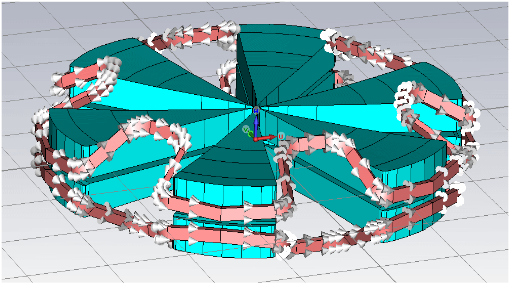}
\caption{\label{AIMA} Layout of the six sector single stage AIMA with reverse valley B-field \cite{Mandrillon:Cyclotrons2016-FRA01}.}
\end{minipage}\hspace{20pc}%
\end{center}
\end{figure}

Unlike the DAE$\delta$ALUS cyclotron, the AIMA proposal is a single stage machine. The latter, discussed as well in section \ref{section:single-stage}, is a single stage six-sector H$_2^+$ cyclotron employing three ion sources that inject three
independent beams in order to increase its reliability. One key feature of the AIMA proposal is the use of superconducting coils with enough reversal field in the valleys as illustrated in fig \ref{AIMA}. This has the merit to increase the magnetic flutter thereby enhancing the vertical focusing at the highest energies and simplifying the extraction path.  

Despite the advantages of molecular H$_2^+$ cyclotrons, some remaining challenges include the stripping foil lifetime and the vacuum level required. Probably the most challenging aspect is the existence of several long lived vibrational states. For this reason, R\&D effort is needed on the ion source to deliver the molecular beam almost free of the highest vibrational states.

In summary, several studies focused on the design of new cyclotrons have demonstrated the possibility to build such a machine with the parameters requested for ADSR. 
However, the construction of a prototype to be finalized for research could be a solution to investigate the technical problems and to validate the viability of cyclotrons/FFAs as drivers for ADSR applications. In particular, new studies have shown that the reliability can be increased and also the cost of the new cyclotrons can be significantly reduced.
However, more dedicated studies are needed to demonstrate that FFAs can be optimized to be fault-tolerant.

\chapter{Novel Concepts}

\chapterauthor[1,2]{H. Okuno}
\chapterauthor[1]{D. Winklehner}
\chapterauthor[3]{J.B. Lagrange}
\chapterauthor[3]{G. D'Agostino}
\chapterauthor[3]{L.H. Waites}
\chapterauthor[3]{H. Barnard}
\chapterauthor[3]{L. Zhang}
\\
\begin{affils}
  \chapteraffil[1]{Editor}
  \chapteraffil[2]{Convener}
  \chapteraffil[3]{Speaker}
\end{affils}

\section{Introduction}
This section describes novel concepts to realize high-power cyclotrons and FFAs.
As was discussed in Chapter \ref{sec:limitations}, the main challenges are 
(1) increased space charge and (2) high beam losses. 
Space-charge leads to beam growth, tune depression,
and difficulties keeping the beam focused, which ultimately again leads to beam 
losses and also potentially poor beam quality.

Before the workshop, we identified novel approaches and invited speakers.
These novel concepts aim to make cyclotrons more compact, improve the stability of 
beam dynamics, simplify the design, improve injection and beam capture, and reduce 
beam losses during extraction.
Not all topics could be covered in detailed presentations.
The five presentations given were (in order of the workshop timetable):
A design for a vertical excursion Fixed Field Accelerator (vFFA), 
easing the design of the electromagnets compared to other FFAs; 
Automating the extraction process without septum in the Innovatron project;
Directly injecting into a compact cyclotron through 
an RFQ; Changing the shape of the inflector (the device bringing the beam into 
the cyclotron) to improve the injected beam quality; and replacing the
usual electrostatic inflector with a magnetic device.
These topics will be discussed in Section \ref{sec:novel_talks}.
Topics that were not covered in presentations, but that we feel are important 
novelties, are: Vortex Motion (VM); Single-stage $\sim$1~GeV machines; And
multiple injection ports. They are summarized in the 
remainder of this introduction.


\newpage
\subsection{Vortex Motion\label{vortex}}
With sufficiently high currents, the space charge in a bunch generates an outward electric force. In combination with the
external focusing forces from the (isochronous) cyclotron,
this causes a \emph{longitudinal-radial coupling} or 
\emph{vortex motion}~\cite{baumgarten:vortex1, baumgarten:vortex2, stammbach1996feasibility,kleeven2016some}.
Experiencing vortex motion, the beam ``curls up'' into
a stationary, almost round (in the radial-longitudinal trace
space) distribution and remains stable there.
A simple, intuitive picture would be that of an 
$\vec{E}\times\vec{B}$ drift, where particles gain additional velocity components depending on the orientation of the 
electric (space charge) and magnetic (cyclotron) fields
acting on them.
In the case of vortex motion, this new velocity component 
is always tangential to the bunch boundary and thus leads
to a motion akin to a hurricane, or \emph{vortex}.

\begin{figure}[tb]
\centering
\includegraphics[width=0.8\textwidth]{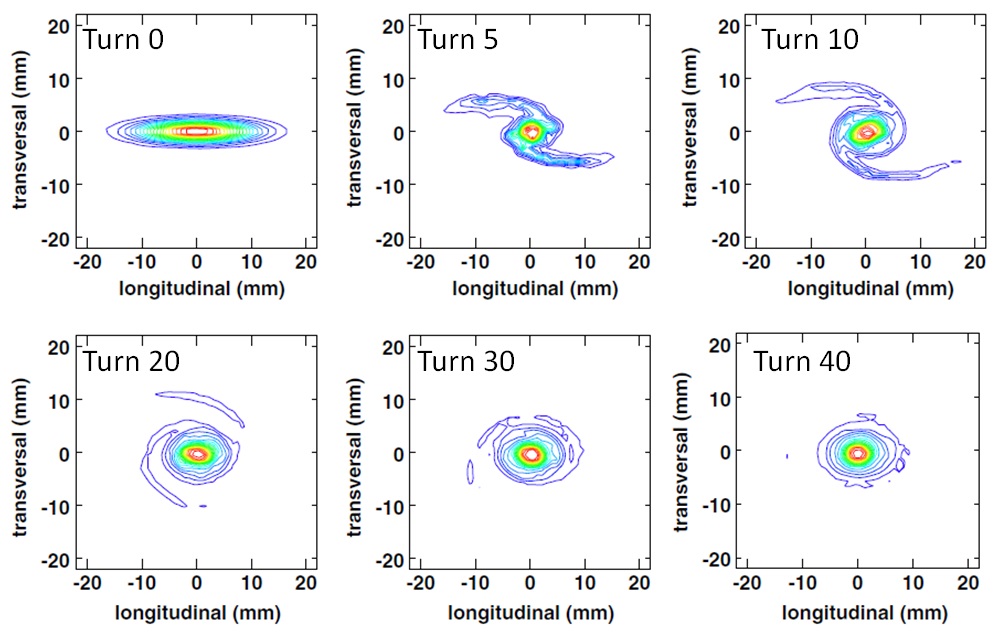}
\caption{\footnotesize Simulations showing the beam profile at various turns within the PSI Injector II cyclotron. Results show the effects of vortex motion on a high intensity beam. From \cite{yang:vortex}. \label{psi_vortex}}
\end{figure}

This effect was first observed at the PSI injector II cyclotron and subsequently simulated accurately using OPAL~\cite{adelmann2008opal, yang:vortex}.
Example results can be seen in Fig.~\ref{psi_vortex}.

While the vortex motion effect was originally discovered at the PSI Injector II cyclotron, no cyclotron has been designed so far to specifically utilize this effect. 
The IsoDAR cyclotron is the first to do so.
Careful tuning of the magnetic field, the vertical size of the beam, the shape of the RF cavities in the central region was
necessary. In addition, optimized collimator placement
in the central region removes beam halo, leading to a 
clean matched distribution~\cite{winklehner2021order}.

\begin{figure}[tb]
\centering
\includegraphics[width=0.8\textwidth]
                {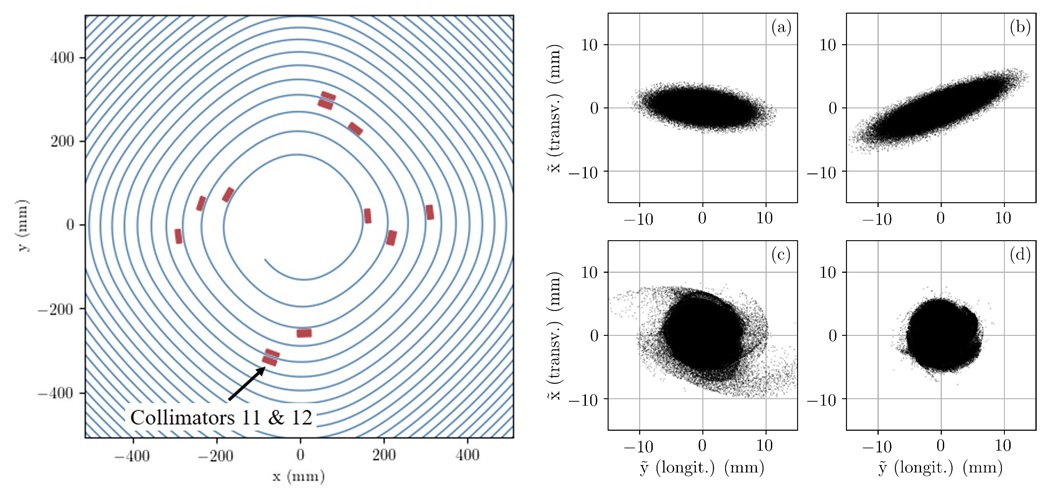}
\caption{\footnotesize (Left) Placement of collimators within the IsoDAR central region optimized to provide the most symmetric beam profile. Removing beam in the low energy central region prevents damage as well as material activation. (Right) Simulations showing the beam profile at turn 7 within the IsoDAR cyclotron under different conditions. (a) Shows the initial beam profile at injection. (b) Shows a simulation which does not include collimators or space charge, effective current=0. (c) Shows a simulation that has accurate space charge to stimulate the vortex motion effect, but does not include collimators in the central region. (d) Shows a simulation that has accurate space charge to stimulate the vortex motion effect,and does include collimators in the central region.
From \cite{winklehner2021order}.\label{isodar_vortex}}
\end{figure}

The round distribution of the beam provides higher turn separation at extraction. Lower levels of overlap in the final turns of the cyclotron are critical for high-power extraction, as any beam overlap at the final two turns will lead to beam power being deposited on the delicate septum which is used for extraction. Using collimators together with vortex motion, the IsoDAR collaboration has been able to show that power on the extraction septum is a factor 2 below the safety limits set
by PSI~\cite{winklehner2021order}.

In addition, in the case of extracting \htp, it is possible to further protect the septum by placing a narrow stripping foil that shadows just the septum. This will remove the molecular electron from the \htp, changing the charge to mass ratio and trajectory of the beam. This will allow for a separate, low power proton beam to be extracted, and negligible power to be deposited on the septum, a crucial issue for high power cyclotrons.

Vortex motion will play in important role in the future design of the highest 
current cyclotrons and it will be necesssary to refine both the analytical understanding
and the simulation tools. The IsoDAR collaboration together with PSI, and 
the OPAL team are undertaking an effort in this directon.

\subsection{Single-Stage 1~GeV Machines}
\label{section:single-stage}
\begin{figure}[t!]
\begin{center}
\begin{minipage}{20pc}
\includegraphics[width=20pc]{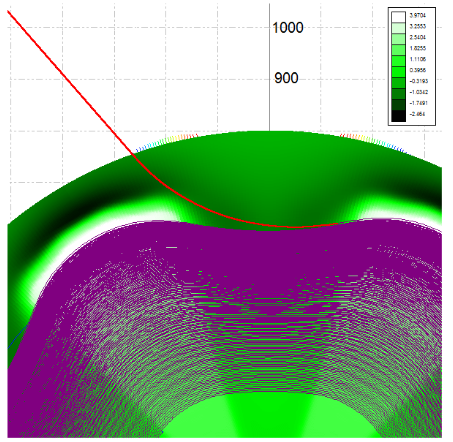}
\caption{\label{AIMA-field} Layout of the extraction path of stripped \htp in the 
         reversed valley field. From~\cite{Mandrillon:Cyclotrons2016-FRA01}.}
\end{minipage}\hspace{20pc}%
\end{center}
\end{figure}

The TRIUMF cyclotron is one of the first concepts of single stage acceleration that exploited the negative H- ions to extract the beam at a final energy of 520 MeV using the stripping extraction method. Due to the electromagnetic stripping effect, the magnetic field shall be low and this sets a major constraint on the size of the machine. The use of \htp molecules provides a remedy to such a problem and is an effective approach to extract the beam even when the turns do overlap. A single stage machine is attractive for several reasons: First, by stripping the \htp in the reversed valley field, the extraction path could be simplified while enhancing the vertical focusing, as proposed for the AIMA cyclotron, shown in Fig.~\ref{AIMA}~\cite{Mandrillon2017SINGLESC} and sketched in fig \ref{AIMA-field}. In addition, the cost of the machine as well as that of the building can be reduced (although for less flexibility). Furthermore, it is possible to use several ion sources in order to mitigate the problem of beam trips and that of the ion source maintenance as discussed in the next section.
Among the challenges is the need for a high quality vacuum, the beam losses due to the dissociation of the vibrational states of the \htp molecule, and the complex shape to obtain the reverse field in the valleys as well as the size of the coils.

\subsection{Multi-port injection}
Multi-port injection is utilized as a means to mitigate the space charge effects which dominate in the first stage of the acceleration regime. This has also the advantage of improving the reliability of the machine, a criteria deemded essential for ADSR. For instance, the AIMA concept foresees the injection of three independent beams of \htp at three different azimuthal positions rotated by $120^{\circ}$ relative to one another. All three beams shall lie in the same median plane of the accelerator and are provided by three low energy injection lines fed by three independent ion sources. It is noteworthy to mention that such an approach is non-trivial to implement for single turn extraction machines given the difficulty to achieve enough turn separation at extraction with three simultaneously circulating beams. Besides, given the interaction between the neighbouring turns, it is important to assess the impact of a beam trip from one of the ion sources on the remaining circulating beams. This is crucial to demonstrate the improved reliability aspect.

\section{Presentations\label{sec:novel_talks}}
\subsection{Talk 1: Development of vertical excursion FFA}
Speaker: J.B. Lagrange, ISIS, RAL, STFC

\subsubsection{Introduction}
The origin of the idea of a vertical excursion Fixed-Field Alternating Gradient accelerator (vFFA) is not new~\cite{Ohkawa1,Teichmann}, but little development was done until recently~\cite{Brooks, Machida}.
In a vFFA, the beams move vertically when accelerated, which has several advantages: in addition to the benefits of the horizontal type of FFA (longitudinal flexibility, sustainability, reliability), since the path length over the whole momentum range is constant, the momentum compaction factor is zero for all orders. For the same reason, isochronism can be achieved for ultra-relativistic energies.

\subsubsection{Beam Dynamics}
To keep the transverse tunes independent of momentum, the magnetic field in a vFFA focusing element satisfies
\begin{equation}
\label{eq1}
\begin{aligned}
B_{x}\left(x,y,z \right)&=B_0\exp\big(m(y-y_0)\big)\sum_{i=0}^Nb_{xi}\left(z\right)x^i,\\
B_{y}\left(x,y,z \right)&=B_0\exp\big(m(y-y_0)\big)\sum_{i=0}^Nb_{yi}\left(z\right)x^i,\\
B_{z}\left(x,y,z \right)&=B_0\exp\big(m(y-y_0)\big)\sum_{i=0}^Nb_{zi}\left(z\right)x^i.
\end{aligned}
\end{equation}
where the $x$-axis is horizontal, the $y$-axis is vertical and the $z$-axis is in the longitudinal direction. $y_0$ is the reference position in the vertical coordinate, $B_{0}$ is the magnetic field at the reference position and $m$ is the normalised field gradient defined as $m=\left( 1/B \right) \left( \partial B/\partial y\right)$. Maxwell's laws can be used to derive the recursive relations between the coefficients $b_{xi}$, $b_{yi}$ and $b_{zi}$.

The expansion of the field in Eq.~\ref{eq1} shows an alternance of skew and normal components. In addition to the skew quadrupole component in the middle of the lattice magnets, there is a potentially strong solenoid field at both ends of the magnet which introduces additional coupling between the horizontal and vertical motions. This makes the modelling of the fringe fields critical in the lattice design of small machines. While the tunes can be obtained as arguments of the conjugate pairs of complex eigenvalues of the computed transverse transfer matrix, the beta-functions follow the Willeke-Ripken procedure~\cite{ripken} to evaluate the beam enveloppe.

To validate the concept of vFFA, a prototype machine is planned at Rutherford Lab in the UK. It will accelerate protons from 3~MeV to 12~MeV, with the ISIS Front End Test Stand (FETS) as an injector. The parameters of the lattice are presented in Table~\ref{tab_fets}.
\begin{table}[ht]
\centering
\caption{\label{tab_fets}%
Main parameters of prototype vFFA lattice.}
\begin{tabular}{ll}
\hline\hline
 parameter & value \\
\hline
Energy & 3 to 12 MeV\\
Repetition & 100 Hz \\
Number of protons per pulse & 3.4 $\times$ $10^{11}$\\
Focusing & FDF triplet\\
Circumference & 28 m\\
Number of cell & 10\\
Total cell length & 2.8 m\\
$B_d$ and $B_f$ magnet core length (M) & 0.50 m\\
Straight length & 1.24 m\\
Distance between $B_d$ centre and $B_f$ centre & 0.53 m\\
Horizontal displacement between $B_d$ and $B_f$ & $\pm$ 0 mm\\
Fringe field parameter (L) & 0.15 m\\
$B_d$/$B_f$ radio (nominal) & 1.15\\
m-value (nominal) & 1.31\\
Orbit excursion & 0.53 m\\
Nominal tune ($q_u$, $q_v$) & 0.756555 / 0.120023\\
Dynamic aperture (normalised) & 60 $\pi$ mm / 70 $\pi$ mm\\
Nominal 100\% emittance (normalised) & 10 $\pi$ mm\\
\hline\hline
\end{tabular}
\end{table}

\subsubsection{Magnet Design}
The maximum magnetic field of a vFFA is typically higher than in a horizontal FFA for a similar application, so using superconducting technology is a natural way to overcome this drawback. Furthermore, it is in line of a more sustainable machine since the power consumption while using it would be lower. A coil dominated magnet would also potentially have more adjusting knobs than a pole shape magnet. The shape of the coil configuration of such a magnet is evaluated analytically based on reversed Biot-Savart principle. A prototype is planned in the coming months at Rutherford Lab. The parameters of the prototype are presented in Table~\ref{tab_proto}.

\begin{table}[ht]
\centering
\caption{\label{tab_proto}Parameters of magnet prototype}
\begin{tabular}{lcr}
\hline\hline
Parameter & value \\
\hline
magnet length & 1.0\,m\\
magnet height & 2.3\,m\\
vertical good field region & 0.6\,m\\
$m$-value & 1.3\,m$^{-1}$\\
\hline\hline
\end{tabular}
\end{table}

\subsubsection{Conclusion}
Thanks to their unique features, vFFAs have been gathering a growing interest recently. Their coupling optics makes it challenging to design. A test ring to demonstrate experimentally the features of the VFFA, called FETS-FFA is planned at Rutherford Lab. A magnet prototype is also investigated to develop a design and manufacturing process.

\subsection{Talk 2: ``INNOVATRON: An innovative industrial high-intensity 
         cyclotron for large-scale production of medical radioisotopes''}
Speaker: G. D'Agostino, IBA

\subsubsection{Introduction}
A research project is ongoing at IBA to study an innovative compact high-intensity self-extracting cyclotron. The project, named InnovaTron, has received funding from the EU H2020 MSCA programme. In the self-extracting cyclotron, proton beams are extracted without any active device. A prototype cyclotron was built by IBA in 2001. Proton currents up to 2 mA were extracted from it. InnovaTron aims at improving the magnet design and the beam optics of the self-extracting cyclotron for the acceleration of high-intensity proton beams up to 5 mA or more to be used for large-scale industrial applications. An overview on the InnovaTron project will be presented together with the first simulation results including space charge. 

Cyclotrons are extensively used for radioisotope production. The self-extracting cyclotron has unconventional features with respect to the existing commercial machines used for this purpose. One of the main features is that the beam from this cyclotron is extracted without using an extraction device (self-extraction). A special shaping of the cyclotron magnetic field and the creation of large turn-separation are used for achieving high extraction efficiency and consequently high extracted beam current \cite{paper1}.\\ 
The proof-of-principle of self-extraction has been already demonstrated by extracting proton currents close to 2 mA from the IBA prototype cyclotron in 2001 \cite{paper2}. The EU InnovaTron project is currently ongoing at IBA for improving the concept of self-extraction in low and medium energy cyclotrons. These are the main goals of the project: i) high proton currents up to 5 mA or more, ii) high extraction efficiency ($>$95\%), iii) reasonable quality of the extracted beam \cite{paper1}. The project opens a new way for large-scale production of medical radioisotopes, such as $^{99m}Tc$ or new emerging PET radioisotopes \cite{paper3,paper4}. 

\subsubsection{Simulation Strategies}\label{intro1}
The approach used to reach the goals of the InnovaTron project consists in an iterative process of optimization of the cyclotron subsystems and of the full integrated cyclotron design. The main tasks of the project are: i) 3D Finite Element (FE) modelling and optimization of the magnet, central region and gradient corrector placed in the extraction region; ii) study of turn separation at extraction, iii) space charge simulations. 
FE modelling tools, such as OPERA, as well as precise 3D beam tracking in the simulated cyclotron electric and magnetic fields are used. The 3D beam dynamics studies have performed with AOC, the IBA’s in-house tracking code used for designing medical and industrial accelerators \cite{paper5}.
Parametrized tools to generate FE models of the cyclotron components have been developed as well as a C code for automated optimization of cyclotron settings by 3D beam tracking from injection up to extraction that maximize the beam properties, such as extraction efficiency and beam quality.

\subsubsection{Space charge simulations}
\begin{figure}[htb]
    \centering
    \includegraphics[width=.4\textwidth]
                {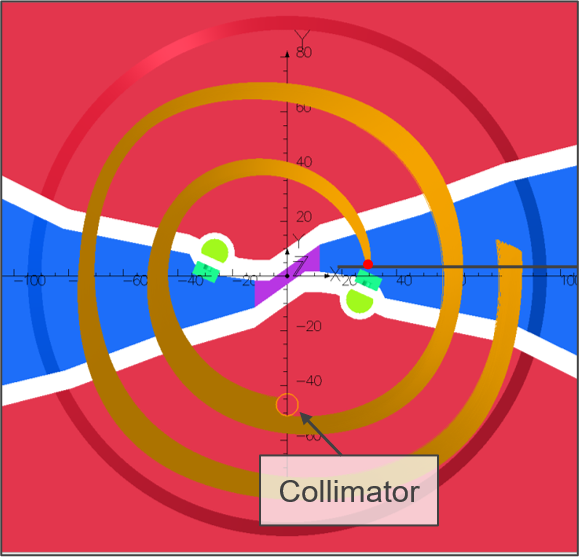}
    \caption{\footnotesize A simulated well-centered beam including space 
             charge in the 3D cyclotron central region model. 
             The bunch starts its motion at the position indicated
             by the red dot in the figure. The black line is the
             observation line of the bunch shape up to turn 20
             plotted in Fig. \ref{ref:figure2}. \label{ref:figure1}}
\end{figure}

A major consideration for achieving high beam current is space charge in the cyclotron center. It makes more difficult to contain and prevent beam losses in the cyclotron center where the magnetic vertical focusing is very weak. Therefore, space charge cannot be neglected in 3D beam tracking simulations when a high beam current is injected in the machine. 
Figure \ref{ref:figure1} shows the beam accelerated in the central region including space charge. A beam current of 5 mA has been assumed. The bunch at the starting tracking position (after the first accelerating gap) has been obtained by beam tracking from the ion source (no space charge) and by processing the beam properties at a chosen time step integration. The red dot in Fig. \ref{ref:figure1} indicates the bunch starting tracking position. A collimator has been placed in the cyclotron center to ensure a well-centered accelerated beam by removing particles with not very good orbit centering.

\begin{figure}[tb]
    \centering
    \includegraphics[width=1.0\textwidth]
	                {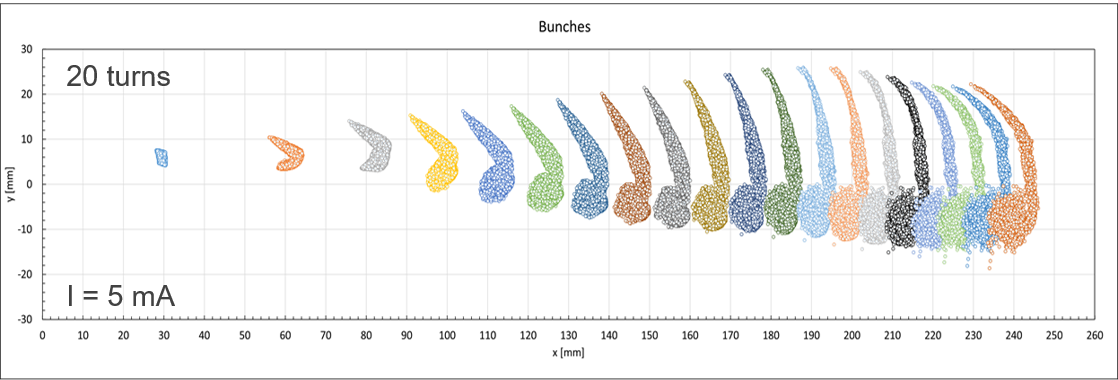}
    \caption{\footnotesize
             Radial-vertical shape of the bunch up to turn 20 
	         along the observation line indicated in
	         Fig.~\ref{ref:figure1}.\label{ref:figure2}}
\end{figure}

Space charge effects induce a vortex motion and an increase of energy spread in the bunch \cite{baumgarten:vortex1}. Figure \ref{ref:figure2} shows the shape of the bunch up to turn 20 along the observation line indicated in Fig. \ref{ref:figure1}. The vortex motion results in a round beam with a more and more intense core turn by turn. Further study is ongoing to see if the beam tail that is remaining can be reduced with collimators in the cyclotron centre.\\ Space charge forces start to act during the process of bunch creation in the first accelerating gap. Furthermore, in simulations of beam tracking in a cyclotron with an internal ion source, the beam phase space at the injection position is not well known. Usually, an educated guess of particle distribution at this position is assumed.
Our recent efforts consist in simulating the plasma meniscus, the beam phase space, and the extracted beam current from the chimney by using the FEM software OPERA. 
Furthermore, we are modelling beam space charge effect in the cyclotron centre, including the bunch formation at the ion source with space charge.

\subsubsection{Conclusion}\label{Assumption_plasma_parameters}

The INNOVATRON project aims at designing a compact high-intensity cyclotron to be used for production of medical radioisotopes. Specific tasks mutually connected are foreseen in the project. Space charge in the cyclotron centre is a major consideration for achieving high beam current. Improvements in 3D beam tracking in a cyclotron including space charge are currently ongoing in the project.

\subsection{Talk 3: RFQ Injection\label{RFQ-DIP}}
Speaker: L.H. Waites, MIT

\subsubsection{Introduction}
One major limitation when going to high power is the
phase acceptance of the cyclotron. In order for particles
to remain bound in the \emph{RF bucket}, they must be
close to the synchronous particle in RF phase (inside the
\emph{separatrix}). There are two aspects to this issue:
(1) Particles that outside of the separatrix might be lost 
in the accelerator, activating the machine or providing
unwanted heat load to a cryogenic system 
(in case of superconducting coils);
(2) Clean turn separation at extraction. Even if
particles are accelerated all the way to the highest energy, 
the beam may suffer energy spread which leads to halo
particles residing in between adjacent turns (i.e. turns
are not well separated) and the septum for electrostatic extraction may be bombarded by high energy particles and
thus activated. H$^-$ machines using stripping extraction 
do not suffer from this issue and can populate about 
~36\degree of RF phase space. Other cyclotrons
accelerating protons, extracting by septum, only 
~10\degree. Vortex motion (cf. Section \ref{vortex})
can alleviate this issue and collimation can be used to
cut away beam that is outside of the phase acceptance.
However, another issue remains:
The ion source and low energy beam transport (LEBT) must 
provide adequate initial currents to be able to cut
the beam outside of the acceptance, often a factor 10 
more than is ultimately accelerated.
This puts unnecessary strain on ion source and LEBT. 
Typically, a (multiharmonic) buncher is placed in 
the LEBT to pre-bunch the beam, compressing more of 
the continuous stream of particles (``DC beam'') into 
a shorter ``bunch''. The efficiency of these bunchers varies,
and high space charge limits the effectiveness due to rapid 
debunching. 

A possible alternative to the LEBT + buncher combination
is using an RFQ embedded axially in the cyclotron yoke to
bunch the beam directly before the spiral inflector
(see Fig.~\ref{isodar_Cyclotron}).



\subsubsection{RFQ Injection}
\begin{figure}[tbh]
\begin{center}
\includegraphics[width=0.5\textwidth]
                {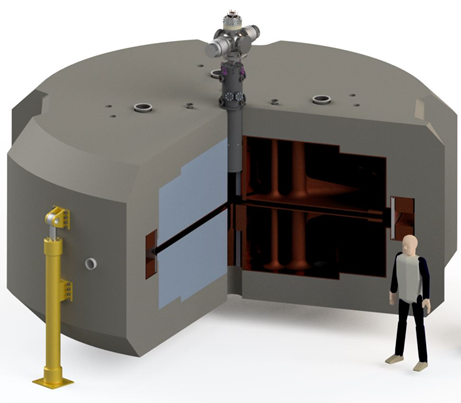}
\caption{\footnotesize
3/4 cut view of a CAD rendering of the IsoDAR cyclotron with RFQ direct injection. \label{isodar_Cyclotron} }
\end{center}
\end{figure}

It is possible to  provide high levels of acceptance and transmission when using a Radio-Frequency Quadrupole (RFQ) to replace the LEBT. An RFQ can be used to accelerate and bunch the beam, and match it to the phase window of the cyclotron. The RFQ direct injection system is designed to inject a bunched and matched beam through a spiral inflector in the center of the cyclotron to be accelerated, as seen in Figure \ref{isodar_Cyclotron}.

The IsoDAR collaboration has developed a design to use this technology for an \htp cyclotron \cite{alonso2021neutrino, winklehner:nima, sangroula2021design, winklehner2016preliminary}.

\begin{figure}[tbh]
\centering
\includegraphics[width=0.8\textwidth]
                {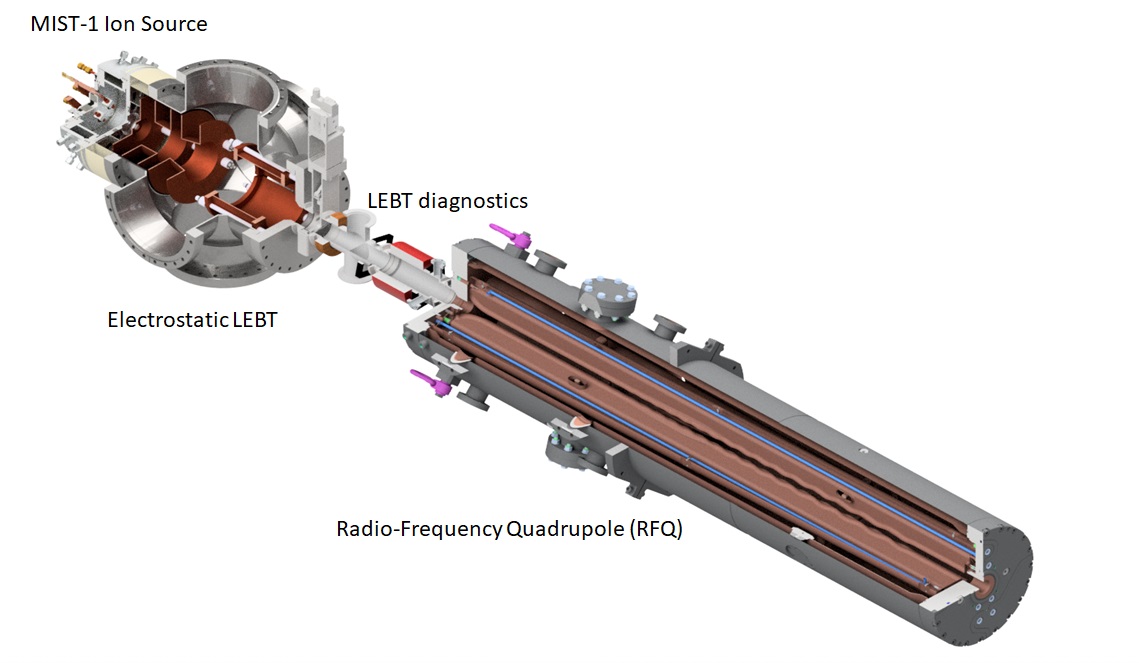}
\caption{\footnotesize
3/4 cut view of the RFQ direct injection system for the
IsoDAR cyclotron. This includes an \htp ion source, 
electrostatic LEBT, diagnostic section, and RFQ. \label{rfq_lebt}}
\end{figure}

The RFQ direct injection system comprises an ion source, a compact electrostatic LEBT, RFQ, and matched central region
with collimators \cite{winklehner2016preliminary}.
The LEBT ensures that the beam from the ion source is properly shaped and matched to the RFQ.
In addition, the LEBT incorporates a chopper~\cite{nath2002beam,waites2020potential}, to
(1) provide lower duty factor beams during commissioning
(lowering the average power while retaining full single bunch charge); And (2) for machine protection. 
Based on simulation work done by the IsoDAR collaboration, this system could provide $>$60\% transmission
from the ion source all the way to beam extracted from the cyclotron~\cite{winklehner2016preliminary, winklehner:nima}.
The design of the injector can be seen in Figure~\ref{rfq_lebt}.

\subsection{Talk 4: Spiral Inflectors}
Speaker: H. Barnard, iThemba Labs

\subsubsection{Introduction}
Injecting the beam into a cyclotron using a Belmont-Pabot spiral inflector poses some difficulties when attempting to match the beam emittance to the cyclotron acceptance. The spiral inflector introduces two unwanted effects: a defocusing of the beam in the vertical direction, and a spreading out of the bunches in the longitudinal direction. The vertical defocusing results in higher losses, mostly during the first few turns of the cyclotron. The longitudinal spreading stems from a coupling between the upstream transverse phase space and the downstream longitudinal position: $(\ell|x), (\ell|x'), (\ell|y), (\ell|y') \neq 0$. This increases the bunch length at the first acceleration gap, counteracting the work of the buncher. 

\subsubsection{Transverse Gradient Inflectors}
Attempts at minimising these problems have resulted in several inflector designs that share the following: The central trajectory and the electric field on the central trajectory correspond to a Belmont-Pabot inflector, but the electrodes are shaped to intentionally produce electric field gradients in the transverse plane. These gradients strongly affect the optics of the inflector, and by selecting them appropriately the optics can be controlled to some degree.

A general description of such a transverse gradient inflector design is obtained by solving the Laplace equation for the potential in the vicinity of the central trajectory. This can be done by expanding the potential up to second order in terms of the transverse coordinates $(u_r,h_r)$ at every point along the path length $s$. It can be shown that the electric fields of such inflectors are fully described by two quadrupole-like functions $Q_1(s)$ and $Q_2(s)$, and the inflector designer is free to select these functions to optimise the optics. At iThemba Labs, an inflector designed using these methods managed to increase the current extracted from the cyclotron by 60\%.    

\subsubsection{Magnetic Inflectors}
\begin{figure}[tb!]
  \centering
  \includegraphics[width=0.4\textwidth]{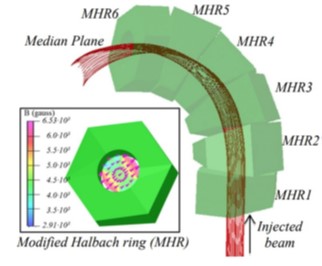}
  \caption{\footnotesize 
           A magnetic inflector comprised of Halbach rings. 
           From~\cite{calabretta_review_2020}.
           \label{fig:mag_inflector}}
\end{figure}

A recent suggestion~\cite{winklehner2014high, calabretta_review_2020} 
has been to replace the electric forces in the spiral inflector 
by identical magnetic forces. The required magnetic fields are produced using 
permanent magnets arranged in a modified Halbach 
ring (see Fig.~\ref{fig:mag_inflector}).
Since the magnetic and electric potentials follow a similar Laplace equation, 
a magnetic spiral inflector can also be described using the two design functions 
$Q_1(s)$ and $Q_2(s)$. To produce these gradients it is suggested that octagonal 
Halbach rings should be used. Initial simulations, where no additional field 
gradients were used, showed that a magnetic spiral inflector had performance similar 
to a standard electric spiral inflector. The main benefits of a magnetic
inflector would be that the risk of electrostatic breakdown is removed, and 
higher injection energies could be achieved.

\subsection{Talk 5: Feasibility Study for the Cylindrically Symmetric Magnetic Inflector}
Speaker: L. Zhang, TRIUMF

\subsubsection{Introduction}
The spiral inflector steers the beam from the bore in the main magnet into the median plane to achieve the axial injection with an external ion source. In a conventional electrostatic inflector, the injection beam energy is limited by the breakdown voltage on the electrodes. While the injection intensity is also limited by the small aperture in the electrostatic inflector. Magnetic inflector is promising to overcome these disadvantages. 

There are two types of magnetic inflectors. One is the passive type which uses the iron in the injection hole to produce the required magnetic field~\cite{kleeven2006injection}. The other is the active one which uses a permanent magnet array~\cite{winklehner2014high}. The passive type is more robust because there is no concern about the magnet degaussing under the high beam loss in the injection hole. But it is only a concept that has no existing design. To demonstrate the technology, we studied the inflection conditions and focal property of the passive magnetic inflector with a cylindrically symmetric structure. 

\subsubsection{Reference Orbit}
In a cylindrically symmetric system. The magnetic vector potential $\operatorname{A}$ only consists of the azimuthal component $\operatorname{A_{\theta}}$. Thus, the Hamiltonian using cylindrical coordinates is written as

\begin{equation}
\begin{aligned}
H=\sqrt{P_{r}^{2} c^{2} + P_{z}^{2} c^{2} + c^{4} m_{0}^{2} + \frac{c^{2} \left(P_{\theta} - q r \operatorname{A_{\theta}}{\left(r,z \right)}\right)^{2}}{r^{2}}}
\end{aligned}
\end{equation}  

Where the canonical momenta are 

\begin{equation}
\begin{aligned}
 P_{r} &= p_r\\ P_{\theta} &=\gamma m_0\theta'r^2+qrA_\theta \\ P_z &=p_z 
\end{aligned}
\end{equation}  

\begin{figure}[tb]
  \centering
  \includegraphics[width=12cm]{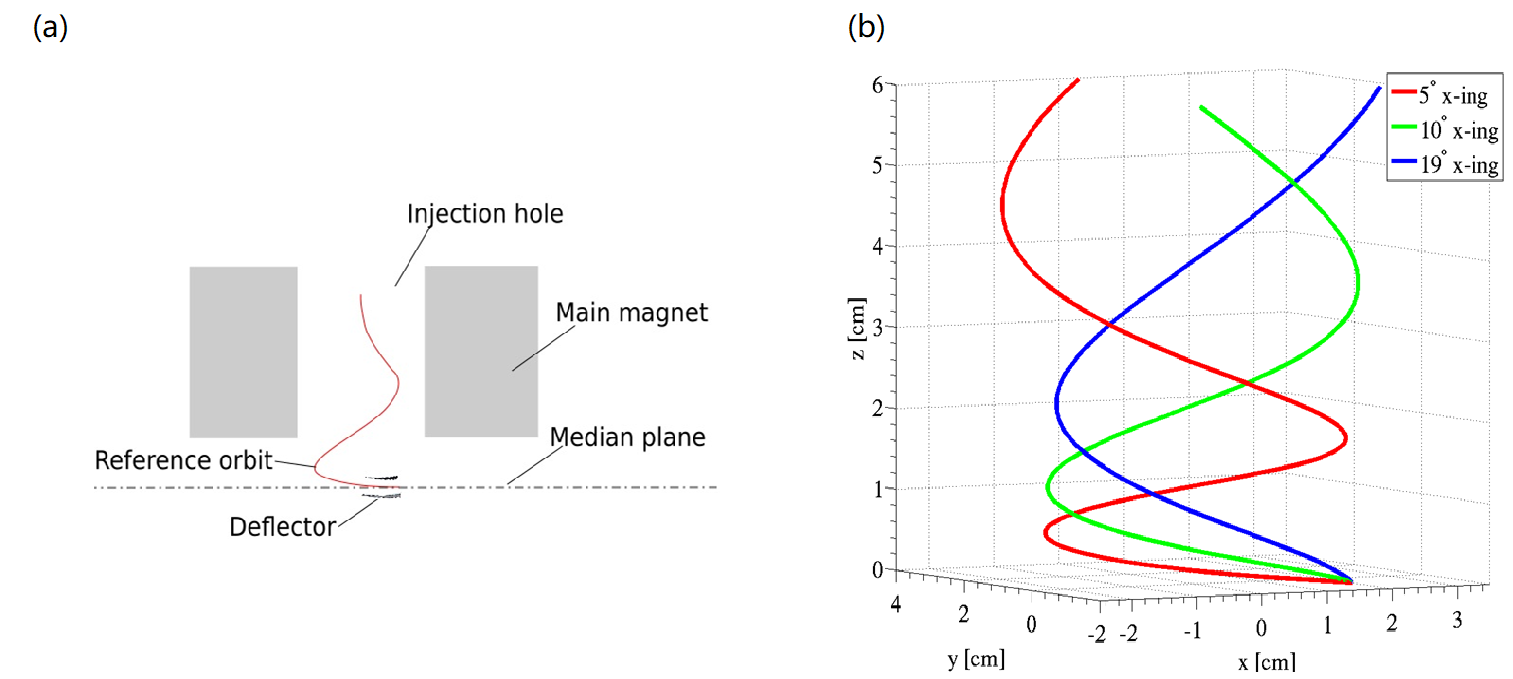}
  \caption{\footnotesize
  Reference orbit in the injection hole.\label{orbit}}
\end{figure}

A vector potential used to define the axial symmetric magnetic field is given as
\begin{equation}
\begin{aligned}
A_{\theta}=\frac{A_1\beta r }{2}-A_2I_1(\beta r)\cos{\beta z}
\end{aligned}
\end{equation} 
Where $\pi/\beta$ is the mirror length, $\beta(A_1+A_2)/(A_1-A_2)$ is the mirror ratio.
We use the TR100 \cite{rao2019} main magnet model as a testbench to study the injection. Figure~\ref{orbit}(a) shows the conceptual model. By tracking the particle reversely from the median plane to the injection point with different Pitch angles, the different reference orbits are shown in figure~\ref{orbit}(b). The single $B_r$ bump field near the median plane could reduce the pitch angle by about 20° from the injection point to the median plane.

\subsubsection{Beam Envelopes}
\begin{figure}[tb]
  \centering
  \includegraphics[width=16cm]{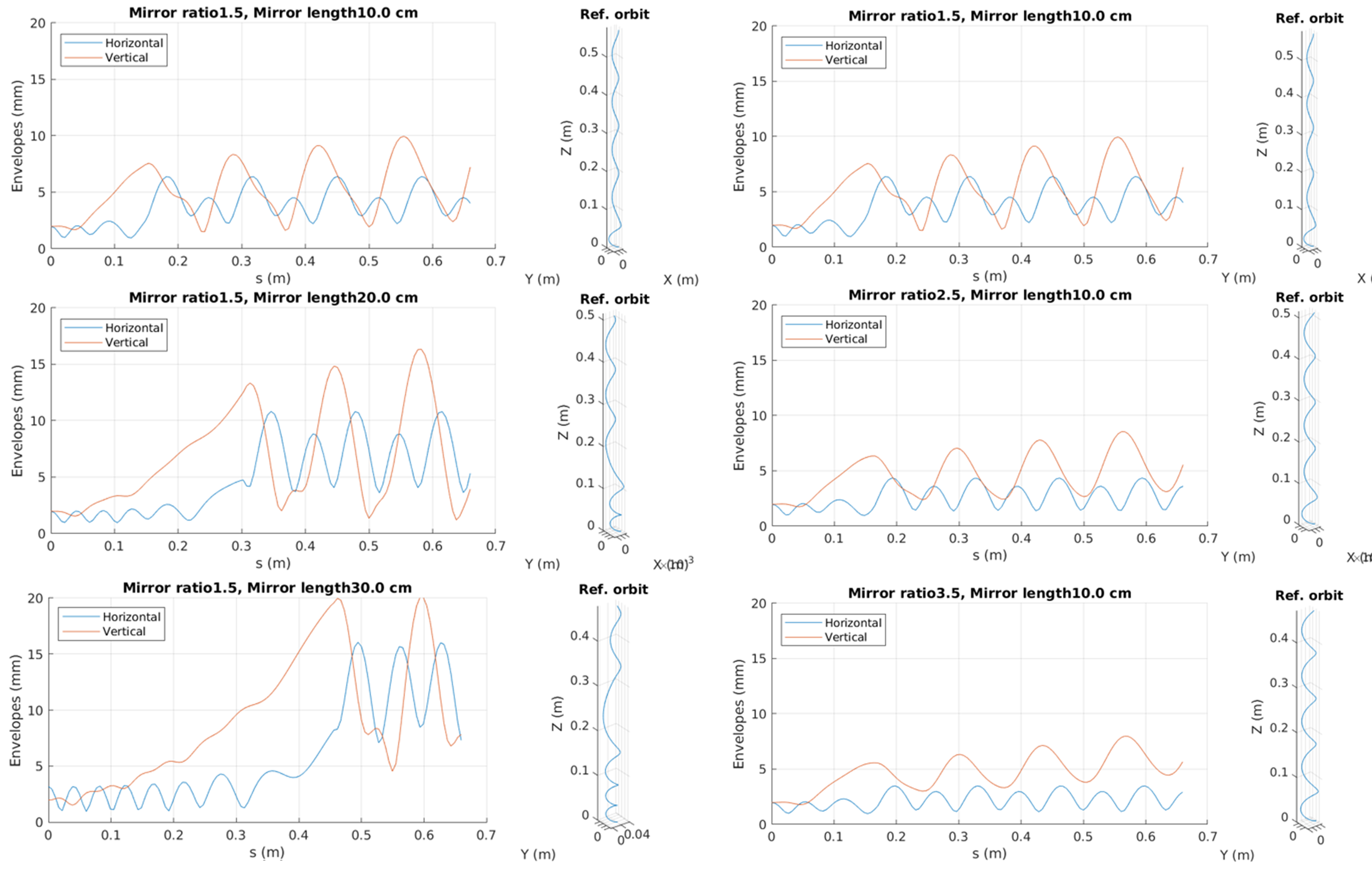}
  \caption{\footnotesize Beam envelopes with different mirror length and mirror ratio. On the right side of each envelopes plot is the reference orbit inside the injection hole. \label{envelope}}
\end{figure}

The beam envelope is studied in the $\alpha-\beta-\gamma$ moving frame. The $\gamma$ direction is the same as the velocity of the reference particle. The $\beta$ direction is perpendicular to the $\gamma$ direction and parallel to the median plane. $\alpha$ direction is perpendicular to both $\beta$ and $\gamma$. Figure~\ref{envelope} shows the horizontal ($\beta$) and vertical ($\alpha$) envelopes with different magnetic field parameters. A proper beam focusing in both directions could be achieved by adjusting the mirror length and the mirror ratio.

\subsubsection{Conclusion}
To maintain the median plane symmetry of the magnet, an electrostatic plate should be placed at the end of the magnetic inflector, which will finally deflect the beam into the median plane with 0 vertical momenta. The envelope study suggests that the beam could be focused both horizontally and vertically by optimizing the mirror ratio and mirror length. Further study for how to design the iron shape that could produce the optimal mirror field should be pursued.

\section{Summary}

Table~\ref{summary} Summarizes the novel concepts for high-power cyclotrons, which appeared in this workshop.
Most concepts are in conceptual design.
R\&D using actual equipment is desired to prove the validity of the concept.

\begin{table}[ht]
\centering
\caption{\label{summary}%
Summary of the novel concepts for high-power cyclotrons.}
\begin{tabular}{llll}
\hline\hline
 Concept&Purpose&Original proposal& Status\\
\hline
"Vortex Motion"&Reduction of beam loss & C. Baumgarten \cite{baumgarten:vortex1}&Conceptual design\\
& at extraction&&\\
\hline
Single-Stage &Reduction of cost&P. Mandrillon\cite{Mandrillon:Cyclotrons2016-FRA01}
& Conceptual design\\
\hline
vFFA& Easy design of Magnets&T. Ohkawa\cite{Ohkawa1}& Conceptual design\\
\hline
Innovatron&Reduction of beam loss&W. Kleeven \cite{paper2}&Simulation study\\
& at extraction&&beyond the POP\\
\hline
RFQ injection&	Improvement of&	R.W. Hamm \cite{RFQdirectinjection}&Conceptual design \\
& injection efficiency&&\\
\hline
Spiral inflectors&Improvement of&	$-$	&Experimental study\\
& injection efficiency&&\\
\hline

Magnetic inflector&	Higher injection energy&W. Kleeven \cite{paper2}&Feasibility study\\
\hline\hline
\end{tabular}
\end{table}

\chapter{Computational Models}

\chapterauthor[1,2]{A. Adelmann}
\chapterauthor[1,3]{T. Planche}
\chapterauthor[1]{P.M. Jung}
\chapterauthor[3]{C. Rogers}
\chapterauthor[3]{P. Calvo}
\\
\begin{affils}
  \chapteraffil[1]{Editor}
  \chapteraffil[2]{Convener}
  \chapteraffil[3]{Speaker}
\end{affils}
This session had 3 presentations, C. Rogers was talking about the {\it Use of a map approach for tracking in FFAs}, P. Calvo gave an account of the {\it Development of the simulation code OPAL}  and T. Planche described the {\it TRIUMF Simulation Tools Status \& Future}. All presentations are avaidable at \url{https://indico.mit.edu/event/150}.

In all high-power accelerators one of the major limitations are particle losses. We distinguish controlled and uncontrolled losses while the later is
most dangerous and unwanted. However, also controlled losses need to be carefully considered, minimized and can not be avoided. A review of available numerical codes can be found in the article of Smirnov~\cite{PhysRevAccelBeams.20.124801}.

\section{Single particle modeling} 
For conventional cyclotrons (and FFAs) the single particle tool box is established 
and many different 
codes and variants exists. For cyclotrons and (horizontal FFAs) the existing 
tools seem to be comfortably and accurate. New machines like vertical FFAs,
currently studies for example at the Rutherford Appleton
Laboratory~\cite{PhysRevAccelBeams.24.021601} require non trivial modifications
to the existing codes. These modifications are on the way for example in the code 
OPAL \cite{OPAL-Manual} and expected to be available in second quarter of 2022.

Recently, in the context of very high field and ultra
compact H$^-$ cyclotrons beam stripping losses of 
ion beams by interactions with residual gas and electromagnetic 
fields are evaluated~\cite{PhysRevAccelBeams.24.090101}.

The beam stripping algorithm, implemented in OPAL, evaluates the interaction of hydrogen 
ions with the residual gas and the electromagnetic fields. In the first case, the cross 
sections of the processes are estimated according to the energy by means of analytical 
functions (see Sec. II-A c\cite{PhysRevAccelBeams.24.090101}). The implementation allows 
the user to set the pressure, temperature, and composition of the residual gas, which 
could be selected for the calculations as either molecular hydrogen (\htp) or dry air 
in the usual proportion. For precise simulations, a two-dimensional pressure field map 
from an external file can be imported into opal, providing more realistic vacuum 
conditions.

Concerning electromagnetic stripping, the electric dissociation lifetime is evaluated through
the theoretical formalism (see Sec. II-B \cite{PhysRevAccelBeams.24.090101}). 
In both instances, the individual probability at each integration step for every 
particle is assessed.

A stochastic evaluation method through an uniformly generated random number is used to evaluate if a physical reaction occurs. In case of interaction, it will be stripped and removed from the beam, or optionally transformed to a secondary heavy particle, compliant with the occurred physical phenomena. In this case, the secondary particle will continue its movement in agreement with the charge-to-mass ratio. Figure 1 summarizes the iterative steps evaluated by the algorithm until the end of the accelerating process, or until the particle is removed from the beam.

\section{Envelope modeling}
A natural extension of single particle modelling, envelope methods are useful for scenarios where the computational time cost of a multi-particle model is too great, but space charge effects are non-negligible. 
Such scenarios include tuning, design, as well as real-time on-line models. 

\section{Multiparticle modeling} 
In general, modelling losses in high intensity accelerators require 3D space-charge and sufficient simulations particles. Recent investigations  \cite{MURALIKRISHNAN2021100094} propose a sparse grid-based adaptive noise reduction strategy for electrostatic particle-in-cell (PIC) simulations. By projecting the charge density onto sparse grids, high-frequency particle noise is reduced and hence an
optimal number of grid points and simulation particle 
can be obtained.

\section{Surrogate Model Construction}
Cheap to evaluate surrogate models have gained a lot of interest lately. 
Statistical \cite{adelmann_2019} or machine learning techniques are used
\cite{info12090351}. These models can for example replace a computational
heavy model in an multi-objective optimization \cite{adelmann-2020-1} 
or in the future be part of the on-line model. 
 
\section{Path forward}
\begin{figure}[tbh]
\begin{center}
\includegraphics[width=0.6\textwidth]{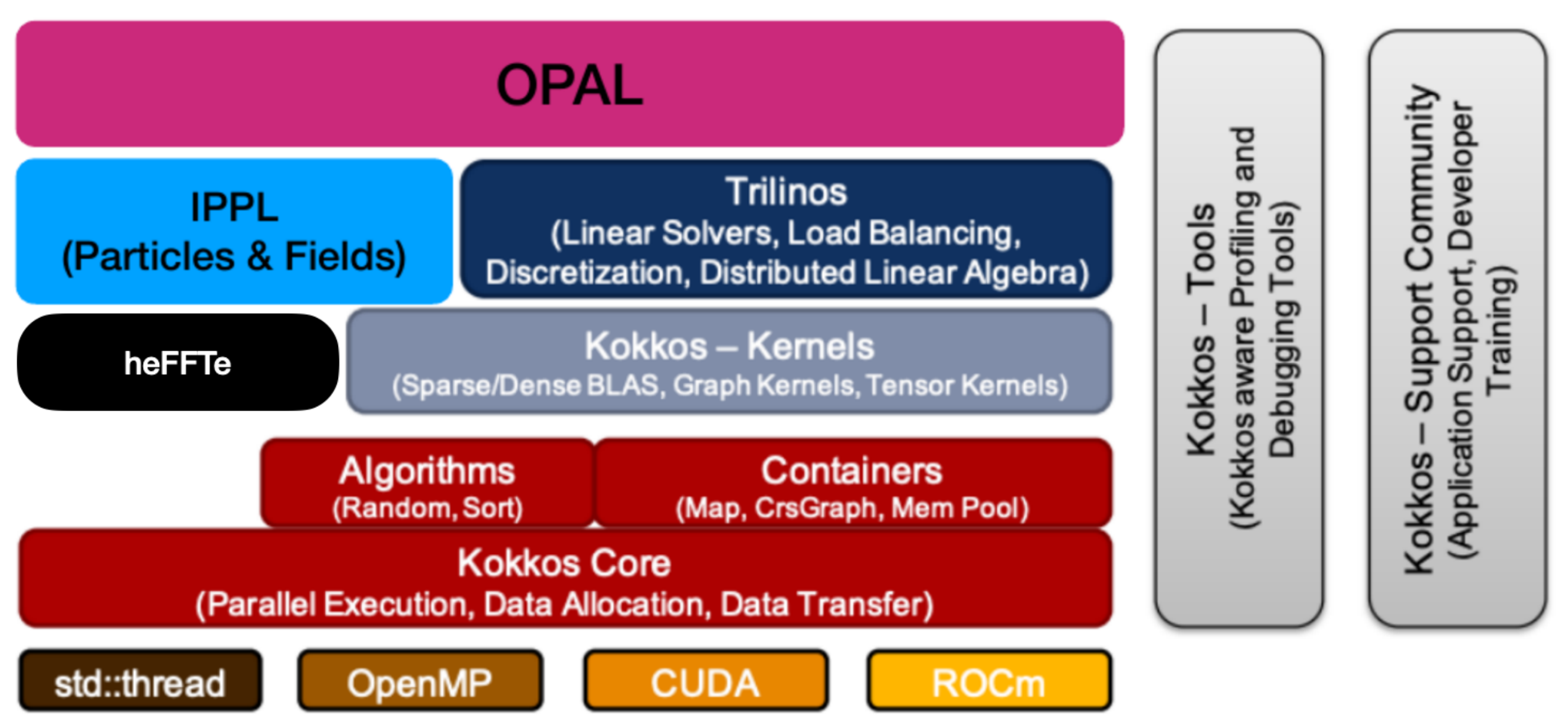}
\caption{Sketch of the new OPAL architecture, based on exiting exascale infrastructure. \label{fig:opalx}}
\end{center}
\end{figure}

While statistical and  machine learning techniques have a lot of potential, high fidelity physics simulations will always be used
to for example produce the training set. In case of high-intensity machines we will need large number of particles and the 
associated fine mesh to solve the PDE in question. It is imperative that we make use of exiting and future high performance infrastructure. A performance portable implementation \cite{frey2021architecture} is of utmost importance. The OPAL collaboration \cite{OPAL-Manual} is in
the progress to completely rewrite the code according to the sketch in Fig.\ref{fig:opalx}. With this new architecture we will be able to make efficiently
use of Exascale-Architecture that will come online soon.

\chapter{Conclusion}

\section{Summary}
Cyclotrons (accelerating hadrons)
have played a major role both in nuclear physics and in particle physics
ever since their invention by E.O. Lawrence in 1930. The relativistic increase of 
inertia limits their maximum energy to $\mathcal{O}$(1)~GeV, thus other accelerator 
types have supplanted them at the \emph{energy frontier}. 
However, due to their ability to
provide cw beams of high current, they are very relevant at the \emph{intensity frontier} -- producing copious amounts of pions, muons, and neutrinos at higher energies and neutrinos from 
isotope decay-at-rest at lower energies. 
For example, the PSI proton facility can deliver up to 
2.4~mA at 590~MeV (a 1.4 MW beam), enabling a vibrant muon program.
IsoDAR is designed to produce 10~mA at 60~MeV (a 600~kW beam), producing neutrinos 
at a rate equivalent to 50 kilocuries.
In addition, cyclotrons have \emph{high societal benefit}
through medical isotope production and energy research.

We found that there have been several breakthroughs in the past years to further 
increase the available beam currents (and thus total delivered power) that make
continuous wave (cw) isochronous cyclotrons the accelerator of choice for many 
high power applications
at energies up to 1~GeV. Key innovations are: Improved injection (through RFQ direct 
injection, transverse gradient inflectors, and magnetic inflectors), improved 
acceleration (utilizing \emph{vortex motion}, single-stage high energy designs,
vertical excursion FFAs), and improved extraction
(through new stripping schemes and by \emph{self-extracting}, using 
built-in magnetic channels). The use of \htp as accelerated ion instead of 
protons or H$^-$ has also received much attention lately. Here, stripping
the electron during extraction or directly after doubles the electrical beam current 
mitigating some of the space charge issues with high current beams in the accelerator.

There are now several projects designing new powerful cyclotrons for particle
physics, medicine, and accelerator driven systems (ADS) for energy research.
These are cost-effective devices with small facility footprint, thus following 
the mantra \emph{better, smaller, cheaper}. Among them, the IsoDAR compact cyclotron
promises a 10~mA cw proton beam at 60~MeV/amu, improving by x4 over PSI injector 2 and
by x10 over commercial cyclotrons for isotope production. A design for a 2~mA 
superconducting cyclotron is underway at TRIUMF, further reducing the footprint.
Several designs (AIMA, DAE$\delta$ALUS, TAMU) are being developed for ADS and
particle phyiscs (CP-violation in the neutrino sector).

Existing simulation tools are addressing the most pressing issues such as controlled and uncontrolled losses with sufficient accuracy and covering complicated new designs. Full statistic resolved start to end simulations of next generation facilities are possible but the time to solution is prohibitive. Surrogate model construction seams to be a viable additional tool, with the goal to decrease time to solution significantly and allowing for example large scale multi-objective optimizations.

\section{Recommendations}
We, the community of particle physicists, particle
accelerator physicists, and funding agencies, should:
\begin{enumerate}[topsep=2pt,itemsep=2pt,parsep=2pt]
\item Recognize the important role cyclotrons are playing in Nuclear- and Particle Physics;
\item Encourage development of this type of accelerator, as an investment with high 
      potential benefits for Particle Physics, as well as outstanding societal value; 
\item Recognize and encourage the high benefit of collaboration with the cyclotron 
      industry.
\item Recognize the opportunities the Exascale era will provide and adjust
      development of beam dynamics simulation tools accordingly.
\item Aim for a close connection of traditional beam dynamics models with 
      (1) machine learning (surrogate models) and 
      (2) feedback (measurements) from the accelerator, as they
      will pave the way to an intelligent accelerator control and on-line optimisation framework.
\end{enumerate}


\clearpage
\appendix
\chapter{Participant List}

\begin{center}
\begin{longtable}{||l l|l||}
\caption{List of registered participants.}\\
\hline\hline
{\bf First} & {\bf Last} & {\bf Institution}\\ 
\hline\hline
\endfirsthead
\caption{List of registered participants (continued).}\\
\hline\hline
{\bf First} & {\bf Last} & {\bf Institution}\\ 
\hline\hline
\endhead
\hline\hline
\endfoot
Andreas & Adelmann & Paul Scherrer Institute \\
Arnau & Albà & Paul Scherrer Institute \\
Jose & Alonso & Massachusetts Institute of Technology \\
Rick & Baartman & TRIUMF \\
Charlotte & Barbier & Oak Ridge National Laboratory \\
\hline
Roger & Barlow & Huddersfield University \\
Hugo & Barnard & iThemba LABS \\
Christian & Baumgarten & Paul Scherrer Institute \\
Yuri & Bylinski & TRIUMF \\
Luciano & Calabretta & INFN - Catania \\
\hline
Pedro & Calvo & CIEMAT, Madrid \\
Grazia & D'Agostino & Ion Beam Applications SA \\
John & Galambos & Oak Ridge National Laboratory \\
Joachim & Grillenberger & Paul Scherrer Institute \\
Hiromitsu & Haba & RIKEN \\
\hline
Malek & Haj Tahar & Paul Scherrer Institute \\
Richard & Johnson & University of British Columbia \\
Carl & Jolly & ISIS, Rutherford Appleton Laboratory \\
Paul & Jung & TRIUMF \\
David & Kelliher & ISIS, Rutherford Appleton Laboratory \\
\hline
Jongwon & Kim & Institute for Basic Science, Daejong, South Korea \\
Daniela & Kiselev & Paul Scherrer Institute \\
Wiel & Kleeven & Ion Beam Applications SA \\
Jean-Baptiste & Lagrange & ISIS, Rutherford Appleton Laboratory \\
Suzanne & Lapi & University of Alabama at Birmingham \\
\hline
Shinji & Machida & STFC Rutherford Appleton Laboratory \\
Mario & Maggiore & INFN-Legnaro \\
Aveen & Mahon & University of Victoria, British Columbia \\
Dmitri & Medvedev & Brookhaven National Laboratory \\
Frank & Meier Aeschbacher & Paul Scherrer Institute \\
\hline
Francois & Meot & Brookhaven National Laboratory \\
Yoshiharu & Mori & Kyoto University \\
Diego & Obradors & CIEMAT, Madrid \\
Hiroki & Okuno & RIKEN \\
Chong Shik & Park & Korea University \\
\hline
Frederique & Pellemoine & Fermilab \\
Thomas & Planche & TRIUMF \\
Christopher & Prior & RAL and Oxford University \\
Danilo & Rifuggiato & INFN Legnaro \\
Chris & Rogers & ISIS, Rutherford Appleton Laboratory (RAL) \\
\hline
Thomas & Ruth & TRIUMF \\
Pranab & Saha & Japan Atomic Energy Agency, J-PARC Center \\
Suzie & Sheehy & University of Oxford \\
Josh & Spitz & University of Michigan \\
Max & Topp-Mugglestone & ISIS, RAL and Oxford University \\
\hline
Loyd & Waites & Massachusetts Institute of Technology \\
Daniel & Winklehner & Massachusetts Institute of Technology \\
Lige & Zhang & TRIUMF \\
Hongwei & Zhao & IMP, Chinese Academy of Sciences \\
Robert & Zwaska & Fermilab \\
\end{longtable}
\end{center}

\chapter{Cyclotron Companies}
\label{sec:companies}
The following table of currently active cyclotron companies
was compiled from the webpages of the IAEA~\cite{iaea_cyclotrons} and
PTCOG~\cite{noauthor_ptcog_nodate}, and
Ref.~\cite{schmor:isotopes}. We distinguish between applications in 
radiotherapy (proton-, helium-, and carbon beams), which typically 
have average beam currents $<10$~$\mu$A, and are not considered ``high-power'',
and medical isotope production, which aim for higher average beam currents,
typically $<1$~mA, which we consider ``high-power''.
We make no claim of completeness.

\begin{center}
\begin{longtable}{||llcl||}
\hline\hline
Full Name & Short Name & Reference & Application\\
\hline
Advanced Cyclotron Systems, Inc. & ACSI & \cite{acsi_web} & Isotopes\\
Best ABT Molecular Imaging, Inc. & Best ABT & \cite{bestabt_web} & Isotopes\\
Best Cyclotron Systems, Inc. & BCSI &\cite{best_web} & Isotopes\\
General Electric Company & GE & \cite{ge_web} & Isotopes\\
Ion Beam Applications & IBA & \cite{iba_web} & Both\\
Mevion Medical Systems & Mevion & \cite{mevion} & Radiotherapy\\
PMB-Alcen & PMB & \cite{imitrace_web} & Isotopes\\
Siemens Healthcare GmbH & Siemens & \cite{siemens_web} & Both\\
Sumitomo Heavy Industries Ltd. & SHI & \cite{sumitomo_web} & Both\\
Thales Group & Thales & \cite{thales_web} & Isotopes\\
Varian Medical Systems, Inc.& Varian & \cite{varian_web} & Radiotherapy\\
\hline\hline
\end{longtable}
\end{center}

\Urlmuskip=0mu plus 1mu
\printbibliography

\end{document}